\documentclass[pdftex,11pt]{article}
\usepackage[utf8]{inputenc}

\usepackage[dvips]{graphicx}
\usepackage{pslatex}

\usepackage{amssymb}
\usepackage{gensymb}
\usepackage{amsmath}
\usepackage{enumitem}

\usepackage{sectsty}
\sectionfont{\fontsize{14}{12}\selectfont}
\subsectionfont{\fontsize{12}{12}\selectfont}

\usepackage[usenames,dvipsnames]{xcolor}
\usepackage{natbib}
\usepackage{hyperref}
\hypersetup{colorlinks,citecolor=MidnightBlue,linkcolor=red,urlcolor=black}

\usepackage[left=1in,right=1in,top=0.9in,bottom=1.1in,footskip=0.5in]{geometry}
\usepackage[onehalfspacing]{setspace}

\usepackage{xurl}

\newcommand{\lsubs}{\textit{L}\textsubscript{\scriptsize{S}}}

\begin{document}

\onehalfspacing

\centerline{\textbf{\Large{Aerosols and tides in the martian tropics during southern}}}
\centerline{\textbf{\Large{hemisphere spring equinox from Mars Climate Sounder data}}}
\vspace{2.0em}
\centerline{L. J. Steele, A. Kleinb\"{o}hl, D. M. Kass and R. W. Zurek}
\vspace{2.0em}
\centerline{Jet Propulsion Laboratory, California Institute of Technology, Pasadena, California, USA.} 
\centerline{Corresponding author: Liam Steele (liam.j.steele@jpl.nasa.gov)}
\vspace{2.0em}
\hrulefill


\vspace{2.0em}
\section*{Key points}

\begin{itemize}[label=\raisebox{0.35ex}{\scriptsize$\bullet$}]
 \item Mars year 29 has large amplitude thermal tides in the tropics during southern hemisphere spring equinox that are not seen in other years.
 \item Data in other years suggests there was a shift in the phase of the diurnal tide to an earlier local time in MY29, due to early dust activity.
 \item Radiative cooling from water ice clouds at night causes increased downwelling and stronger temperature inversions due to adiabatic warming.
\end{itemize}


\newpage
\section*{Abstract}

We analyze Mars Climate Sounder temperature and aerosol data in the tropics to study atmospheric tides and their relation to the dust and water ice distributions. Our results from data covering Mars years (MY) 29--35 reveal that MY29 has large amplitude non-migrating thermal tides during southern hemisphere spring equinox that are not observed at the same local time in any other year. It is the nighttime temperatures that are most perturbed compared to other years, with strong temperature inversions at 35--55 km altitude. Analysis of data at different local times reveals that the temperatures and water ice clouds at 03:45 am in MY29 more closely resemble those at 05:00 am in other years, suggesting there was a shift in the phase of the diurnal tide to an earlier local time. This phase shift, and the large amplitude non-migrating thermal tides, appear to be related to early dust activity. Two early dust storms occurred in MY29 around the time there was upwelling over the tropics, associated with the Hadley circulation, enabling the dust to be transported to higher altitudes where it has a larger radiative influence. 
As well as dust, water ice clouds are also found to influence the tidal structure.
Due to the interaction of non-migrating tides, water ice clouds occur in two discrete longitudinal regions at night. The increased radiative cooling results in increased downwelling above the clouds, leading to increased adiabatic warming and a strengthening the temperature inversions.


\newpage
\section*{Plain Language Summary}

Tides can cause oscillations in Mars' atmosphere, just like they can cause oscillations in the Earth's oceans. The tides on Mars are caused by variations in the heating of the atmosphere by the sun, and they can affect temperatures, wind, pressure, and the transport of aerosols. We investigate these tides using data from the Mars Climate Sounder instrument onboard the Mars Reconnaissance Orbiter spacecraft. We find that while the daytime atmosphere in different years is similar, the nighttime atmosphere in one particular year - Mars year 29 - has a noticeably different structure. We determine that this difference is due to a combination of the effects of both dust and water ice. These aerosols cause regions of localized heating and cooling, which affects the structure of the tides, and the circulation of the atmosphere.


\newpage
\section{Introduction}\label{sec:intro}

Thermal tides are global-scale oscillations of atmospheric temperature, pressure and wind, which are sub-harmonics of a solar day. They are driven by solar heating of the atmosphere and surface, and due to Mars' thinner atmosphere they play a much more significant role than on Earth. They can influence surface pressure and wind, affect the transport of aerosols, and propagate into the upper atmosphere \citep{Zurek1976, WilsonHamilton1996, Forbes2002}. Observations of thermal tides on Mars date back to the 1970s, when data from Mariner 9 and the Viking landers and orbiters were analyzed \citep[e.g.][]{Leovy1973, Zurek1976, Zurek1988, Conrath1976, LeovyZurek1979, Leovy1981, WilsonRichardson2000}. Since this time, orbiters and surface instruments have provided a wealth of data about the tides.

Data from the Rover Environmental Monitoring Station (REMS) on board the Mars Science Laboratory (MSL) are the most complete set of surface data since the Viking landers. MSL is located near the equator within Gale crater, and the tropics are a region with weather dominated by tides \citep{WilsonHamilton1996, HinsonWilson2004, Hinson2008}. REMS measurements show that diurnal and semidiurnal tides in Gale Crater are highly correlated with atmospheric opacity, with enhanced tidal amplitudes observed during the Mars Year (MY) 34 dust storm \citep{Guzewich2016, Guzewich2019, Viudez2019}.

Vertical profiles of global atmospheric temperatures in the lower and middle atmosphere (at altitudes below $\sim$80 km), particularly from the Thermal Emission Spectrometer (TES) and Mars Climate Sounder (MCS) instruments, provide a rich data set with which to analyze thermal tides. Such data, covering multiple Mars years, have revealed the presence of migrating (sun-synchronous) tides, as well as non-migrating tides that can move westward or eastward, with zonal wavenumbers of 1--4  being most prominent \citep{Banfield2000, Banfield2003, Lee2009, Guzewich2012, Kleinbohl2013, Guzewich2014, Wu2015, Wu2017}. Thermal tides have also been detected in the thermosphere, with analysis of accelerometer data from the aerobraking phases of Mars Global Surveyor and Mars Odyssey revealing zonal wavenumber 2--3 patterns in density at altitudes of 100--160 km \citep{Keating1998, Withers2003, Wang2006}. More recently, Mars Atmosphere and Volatile EvolutioN (MAVEN) measurements have been used to detect zonal wavenumber 1--3 structures at altitudes of up to 190 km \citep{Lo2015, England2016, Stevens2017, Jiang2019}. MGS Radio Science measurements have revealed tidal effects up to altitudes of 200 km \citep{Bougher2004, Cahoy2006}, while Mars Reconnaissance Orbiter (MRO) radio tracking data have revealed zonal wavenumber 1--2 patterns in density in the exosphere, at $\sim$250 km \citep{Mazarico2008}.

Numerical simulations and classical tidal theory suggest that the zonal patterns observed in the atmosphere are the result of thermal tides propagating from at or near the surface. These tides are modulated by variations in Mars' surface properties (topography, thermal inertia and albedo), the atmospheric dust loading, water ice clouds, and interactions between planetary waves and tides \citep{Zurek1976, Conrath1976, Joshi2000, ForbesHagan2000, Wilson2002, HinsonWilson2004, ForbesMiyahara2006, MouddenForbes2008a, MouddenForbes2008b, MouddenForbes2010, MouddenForbes2011}. Eastward-propagating Kelvin waves are found to be the dominant component for zonal wavenumbers $> 1$ \citep{Wilson2000, Banfield2003, Withers2011, Guzewich2012}, as these are close to resonance in the martian atmosphere \citep{Zurek1988, WilsonHamilton1996}.

For global temperature data from low polar orbiters, available at two local times 12 hours apart ($\sim$2 am/pm for TES data and $\sim$3 am/pm for MCS in-track data), the standard tidal analysis technique is to perform Fourier analysis of the average and difference fields of the two local times \citep[e.g.][]{Banfield2003, Lee2009, Guzewich2012}. To date, studies utilizing MCS data have only performed Fourier analysis on the in-track measurements, observed in the direction of MRO's orbital movement. MCS also performed cross-track scanning, viewing the limb at 90\degree{} to the left and right of the orbit trajectory, obtaining measurements between 1.5--3 hours earlier and later in local time than the in-track measurements, depending on latitude. \citet{Kleinbohl2013} used the cross-track data to study tides by applying nonlinear least-squares fits to the time series data, but Fourier analysis of the data, and investigations of the aerosol behavior, have not been performed. We perform this additional analysis here, in order to investigate the temporal behavior of thermal tides and their relation to aerosols. In particular, we focus our attention on the tropics, as this region was not analyzed in earlier studies.


\section{Data and analysis method}\label{sec:data}

\begin{figure}[t]
  \begin{center}
  \includegraphics[width=0.47\linewidth]{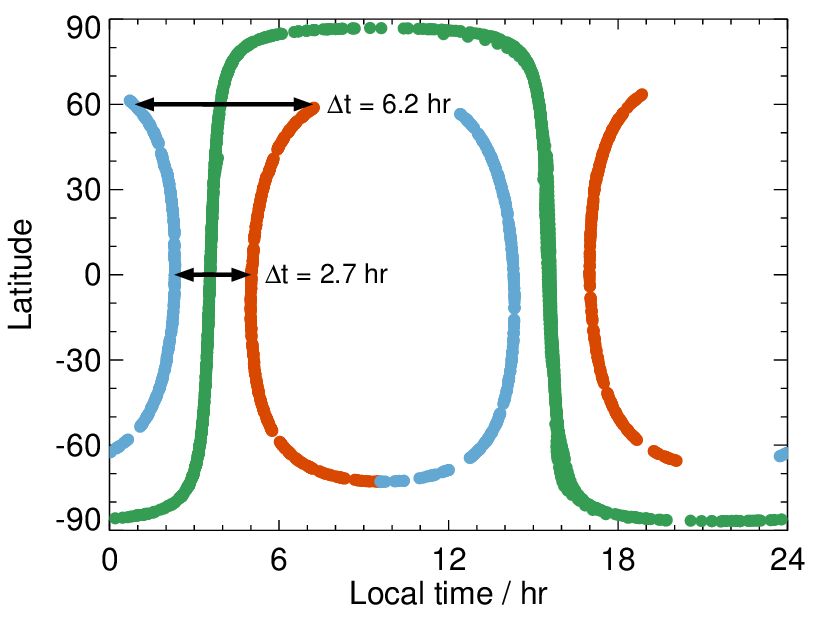}
  \end{center}
  \caption{Local time coverage of MCS in-track (green) and cross-track (blue, red) measurements. Data are from MY30, \lsubs{} = 185--190\degree{}.}
  \label{fig:fig_loct}
\end{figure}

MCS is a nine-channel limb radiometer on board MRO, with 8 mid- and far-infrared (IR) channels and one visible/near-IR channel. Each channel consists of 21 detectors, providing a radiance profile from the surface to $\sim$90 km, with $\sim$5 km resolution \citep{McCleese2007}. The retrieval algorithm inverts the radiance profiles to produce geophysical quantities, including vertical profiles of temperature, dust and water ice extinction \citep{Kleinbohl2009, Kleinbohl2011, Kleinbohl2017}. MCS typically observes the limb in the direction of MRO's orbital movement, producing in-track measurements. However, beginning in MY30, and continuing through MY35, the observation strategy was modified to also view the limb at 90\degree{} to the orbit trajectory, producing cross-track measurements \citep{Kleinbohl2013}. Figure \ref{fig:fig_loct} shows how the local times of the measurements vary with latitude. As the orbit moves from the equator to $\pm$60\degree{} latitude, the time spanned by the earlier (blue) and later (red) cross-track measurements increases from $\sim$2.7 hours to $\sim$6.2 hours. For simplicity, and due to focusing on the tropics, when referring to local times we will reference the equator-crossing local time.

In order to perform Fourier analysis on the MCS temperature retrievals, we first binned the data spatially and temporally on the MCS retrieval pressure levels. We binned the data by 15\degree{} in longitude, 5\degree{} in latitude and 5\degree{} in \lsubs. These values were chosen to ensure the majority of bins were filled with data, but that bins were not too large that potentially interesting spatial and temporal structures were missed. As Fourier analysis requires all longitudinal bins at a given latitude and pressure level to be populated, bins with missing data were filled using cubic spline interpolation. Interpolation was not performed if two consecutive longitudes bins had missing data, or if more than 25\% of the longitudinal bins had missing data. For the period we are interested in (\lsubs{} $\approx$ 180--360\degree{}) there are typically one or two longitudinal bins in the tropics during the daytime that require interpolation. Away from the tropics, and at most locations during the night, interpolation is typically not required.

For the in-track data (green symbols in Figure \ref{fig:fig_loct}), and the cross-track data at earlier and later local times (blue and red symbols in Figure \ref{fig:fig_loct}), binning produced morning (am) and afternoon (pm) temperature fields, from which averages and differences were created via:

\vspace{-1em}
\begin{align}
  T_\mathrm{ave}  &= (T_\mathrm{pm} + T_\mathrm{am})/2, \\
  T_\mathrm{diff} &= (T_\mathrm{pm} - T_\mathrm{am})/2.
\end{align}

For periods with only in-track data this resulted in one $T_\mathrm{ave}$ and one $T_\mathrm{diff}$ field, while for periods with in-track and cross-track data there were three $T_\mathrm{ave}$ and three $T_\mathrm{diff}$ fields. Fourier analysis was then performed, in order to obtain the amplitudes and phases of the tides. This assumes that at a given latitude and pressure level, the temperature variation with longitude, $\lambda$, and local time, $t_\mathrm{LT}$, can be expressed as

\vspace{-1em}
\begin{align}
  T(\lambda,t_\mathrm{LT}) &= {\sum}_{s,\sigma} T_{s,\sigma} \cos[(s-\sigma)\lambda + \sigma t_\mathrm{LT} - \delta_{s,\sigma}] \nonumber \\
                           &= {\sum}_{s,\sigma} T_{s,\sigma} \cos(m\lambda + \sigma t_\mathrm{LT} - \delta_{s,\sigma}),
\end{align}
where $s = 0, \pm1, \pm2...$ is the zonal wavenumber (with positive and negative values denoting westward and eastward propagation respectively), $\sigma = 0, 1, 2...$ is the frequency, $m = |s - \sigma|$ is the satellite-relative wavenumber, $T_{s,\sigma}$ is the amplitude of the tide and $\delta_{s,\sigma}$ is the phase. A stationary wave has $\sigma = 0$, while diurnal and semidiurnal tides have $\sigma = 1$ and $\sigma = 2$, respectively. Fourier analysis of $T_\mathrm{ave}$ and $T_\mathrm{diff}$ provides different information on the tides present. For example, only even (odd) values of $\sigma$ appear in the $T_\mathrm{ave}$ ($T_\mathrm{diff}$) fields. Some thermal tides can propagate vertically in the atmosphere, while others are trapped in the region of excitation. Additionally, thermal tides can be decomposed into components which are symmetric or anti-symmetric about the equator. (\citealp[For more details see][]{Forbes1995, Banfield2003, Lee2009, Guzewich2012, Guzewich2014}.)

When discussing various tidal components we will use `D' and `S' to denote diurnal and semidiurnal tides, and `E' and `W' to denote eastward and westward propagation, with the number following denoting the zonal wavenumber $s$. For example, DE1 represents an eastward-propagating diurnal wavenumber 1 tide, while SW2 denotes a westward-propagating semidiurnal wavenumber 2 tide. Zonally-symmetric tides, which do not propagate eastward or westward and have no zonal wavenumber, are represented by D0 or S0. Tides which propagate westward with the apparent phase speed of the Sun (e.g.\ DW1 and SW2) are called migrating tides, while the others are called non-migrating.


\section{Results}\label{sec:tides}

\subsection{Tide amplitudes}

We first look at the amplitudes from the Fourier analysis of the $T_\mathrm{ave}$ and $T_\mathrm{diff}$ fields. We focus on satellite-relative wavenumbers \textit{m} = 1--3, as these have the largest amplitudes in the MCS data. After analyzing tides from different Mars years, it is clear that the behavior in the tropics in MY29 around southern hemisphere spring equinox (\lsubs{} $\approx$ 160--200\degree{}) is different to other years, and is noticeable for the appearance of tides with amplitudes $> 5$ K. The atmosphere around this time in MY29 was dustier than in other years observed by MCS due to two early local dust storms \citep{WangRichardson2015}. One storm originated in the northern hemisphere at \lsubs{} $\approx$ 143\degree{}, while another originated in the southern hemisphere at \lsubs{} $\approx$ 152\degree{}.

\begin{figure}[t]
  \begin{center}
  \includegraphics[width=1.0\linewidth]{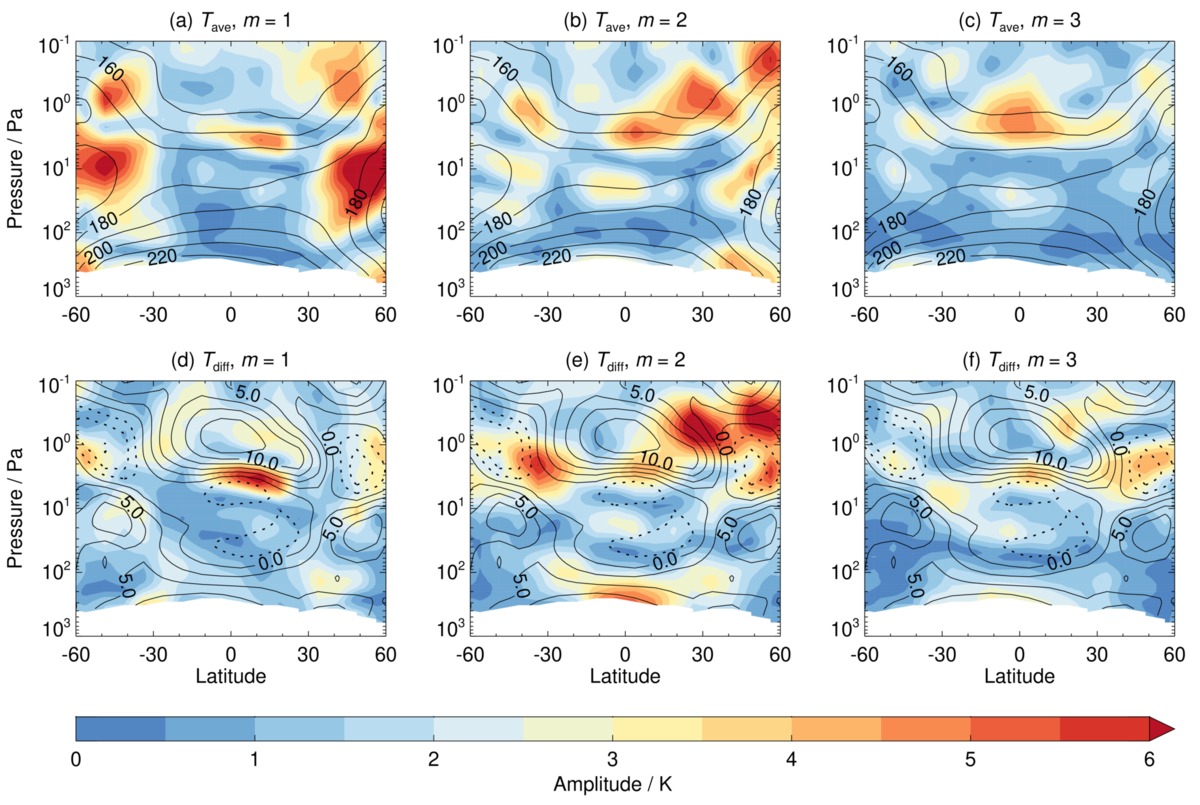}
  \end{center}
  \caption{Amplitudes of the \textit{m} = 1--3 components of the (a--c) $T_\mathrm{ave}$, and (d--f) $T_\mathrm{diff}$ fields, averaged over \lsubs{} = 180--200\degree{}, MY29. Black contours show the \textit{m} = 0 component (10 K intervals for $T_\mathrm{ave}$ and 2.5 K intervals for $T_\mathrm{diff}$), with dashed lines representing negative values. Data are from in-track measurements at a local time of 3:45 am/pm.}
  \label{fig:ampl_my29_in}
\end{figure}

Figure \ref{fig:ampl_my29_in} shows the \textit{m} = 1--3 components of the $T_\mathrm{ave}$ and $T_\mathrm{diff}$ fields for MY29, with the data averaged over \lsubs{} = 180--200\degree{}, which is when the amplitudes in the tropics are at their largest. The black contours overlain on the $T_\mathrm{ave}$ and $T_\mathrm{diff}$ fields show the \textit{m} = 0 components. For $T_\mathrm{ave}$ this component is mostly dominated by the time- and zonal-mean temperature (for which $s = \sigma = 0$), but will also have contributions from the sun-synchronous semidiurnal tide, SW2, and possibly higher even-harmonic sun-synchronous tides. For $T_\mathrm{diff}$ this component is mostly dominated by the sun-synchronous diurnal tide, DW1, with additional contributions from higher odd-harmonic sun-synchronous tides \citep{Zurek1976, Banfield2003, Lee2009, Guzewich2012, Kleinbohl2013}.

The tidal features in the mid- to high- latitudes have been discussed in detail previously \citep{WilsonHamilton1996, Banfield2000, Banfield2003, Lee2009, Guzewich2012} so are not considered further here. Instead we focus on tides in the tropics. In the $T_\mathrm{ave}$ fields (Figure~\ref{fig:ampl_my29_in}a--c), the regions of large amplitude tropical tides are located between $\sim$1--5 Pa. As \textit{m} progresses from 1--3, the vertical extents of these regions increase, and they shift southwards in latitude, becoming centered on the equator. In the $T_\mathrm{diff}$ fields (Figure~\ref{fig:ampl_my29_in}d--f) the large amplitudes all occur at a similar height, centered at $\sim$3 Pa, which corresponds to a region of strong gradients in the diurnal tide (black contours). They are offset north of the equator with relatively small vertical extents, and the amplitudes decrease with increasing $m$. For \textit{m} = 2 and \textit{m} = 3 in both the $T_\mathrm{ave}$ and $T_\mathrm{diff}$ fields there are two weaker tidal signatures below the main ones identified, at $\sim$25 Pa and $\sim$300 Pa, which are centered more on the equator. 

\begin{figure}[t]
  \begin{center}
  \includegraphics[width=1.0\linewidth]{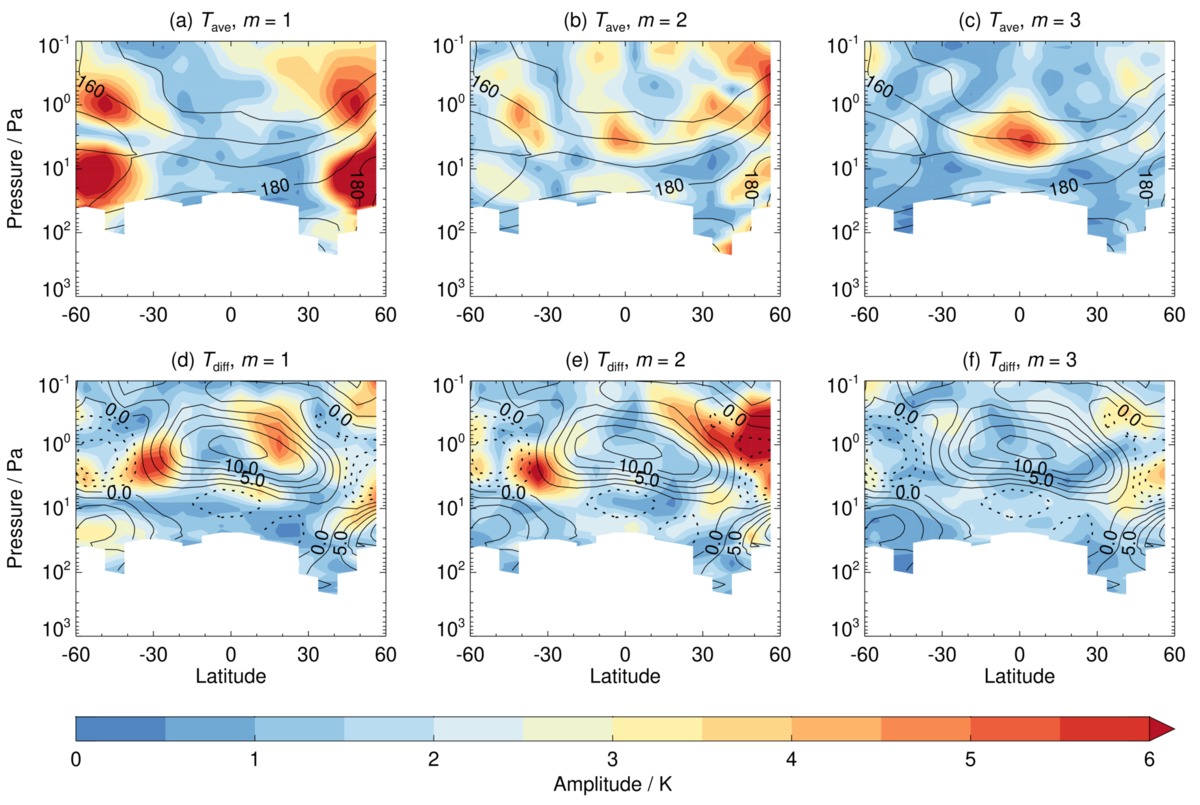}
  \end{center}
  \caption{As Figure \ref{fig:ampl_my29_in}, but from MY30 cross-track measurements at a local time of 5:00 am/pm.}
  \label{fig:ampl_my30_cr}
\end{figure}

Such large-amplitude tropical tides are not seen in the in-track measurements in other Mars years (see Figure S1), but some similar tidal features do appear when looking at cross-track measurements taken $\sim$1.5 hours later in local time. Figure \ref{fig:ampl_my30_cr} shows the \textit{m} = 1--3 components of the $T_\mathrm{ave}$ and $T_\mathrm{diff}$ fields from MY30 cross-track measurements at local times $\sim$5:00 am/pm. MY30 was chosen for comparison as it is representative of the behavior seen in MY31--33 and MY35 in this \lsubs{} period (MY34 experienced a global dust storm around this time, so cannot be compared, \citealp[e.g.][]{Kass2019, Kleinbohl2020}). Large amplitude tropical tides of 4--6 K appear at this later local time in the \textit{m} = 2 and \textit{m} = 3 components of $T_\mathrm{ave}$ (Figure \ref{fig:ampl_my30_cr}b,c), and a smaller amplitude tide of 2--3 K appears in the \textit{m} = 1 component of $T_\mathrm{diff}$ (Figure \ref{fig:ampl_my30_cr}d). These tides appear in similar locations to those seen in the MY29 in-track measurements, with again the $T_\mathrm{ave}$ amplitudes more vertically-extensive, and the $T_\mathrm{diff}$ amplitude located in the region of strong gradients in the diurnal tide.

\begin{figure}[t]
  \begin{center}
  \includegraphics[width=0.8\linewidth]{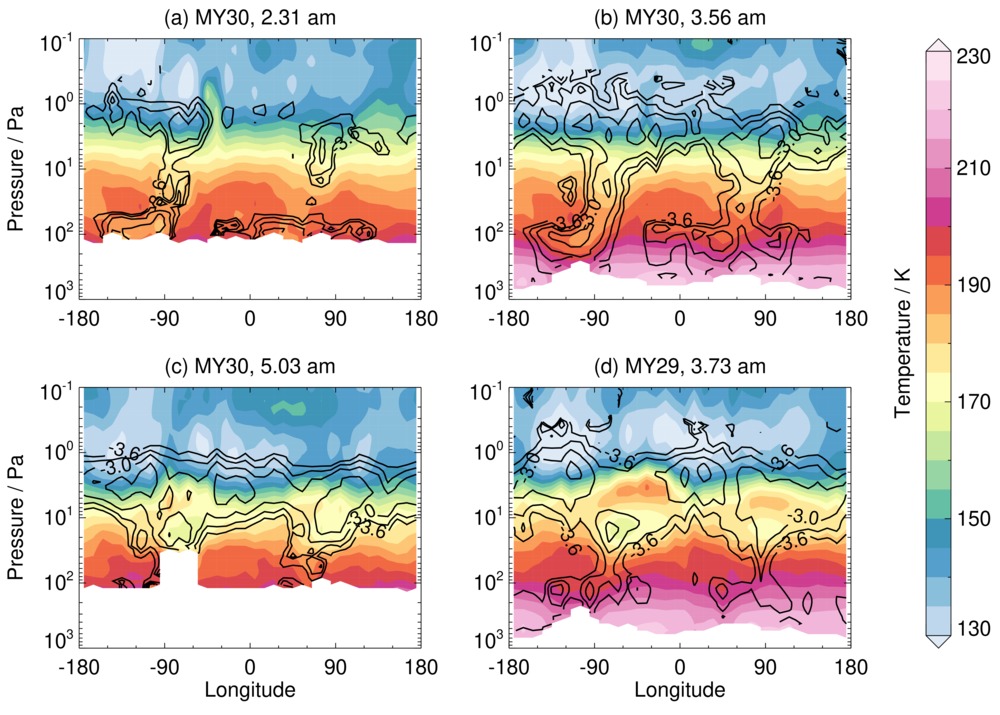}
  \end{center}
  \caption{Temperatures (shaded) and the log of the water ice opacity per km (contours) from (a--c) MY30 in-track and cross-track data at different local times, and (d) MY29 in-track data. Ice opacity contours range from $-$3.6 to $-$2.4, in steps of 0.3. Data are averaged between 0--10\degree{}N and \lsubs{} = 180--200\degree{}. (See Figure S3 for a similar plot of the shaded water ice cloud distribution.)}
  \label{fig:ice_temp_2my_night}
\end{figure}

As the tide amplitudes in the Fourier analysis are derived from daytime and nighttime temperature measurements, we can also look at these two times of day individually. The temperature distributions in the daytime are broadly similar in MY29 and MY30 (see Figure S2), but the nighttime temperature distributions are markedly different in MY29 compared to MY30. Thus, it is the nighttime temperature structure which is responsible for the large amplitude tides seen in the MY29 in-track data, and not in the MY30 in-track data. Figure \ref{fig:ice_temp_2my_night} shows the temperature and water ice cloud distributions, as a function of longitude and altitude, from MY29 data at a local time of 03:45 am, and MY30 data at three different local times (02:20 am, 03:35 am and 05:00 am). Data are averaged between 0--10\degree{}N, as this latitude range contains large amplitudes for all tidal components in MY29. Comparing the in-track data in both years, it is clear that MY29 has prominent temperature inversions at 03:45 am between $\sim$3--20 Pa (Figure \ref{fig:ice_temp_2my_night}d) which are not seen at 03:35 am in MY30 (Figure \ref{fig:ice_temp_2my_night}b). The inversions are located between 15--90\degree{}W and 75--110\degree{}E, with the inversions being strongest in the western hemisphere, where temperatures at 4 Pa are $\sim$30 K warmer than in MY30. However, when looking at a later local time of 05:00 am in MY30 (Figure \ref{fig:ice_temp_2my_night}c), temperature inversions can be seen, and this results in the appearance of the tropical tide signals seen in Figure \ref{fig:ampl_my30_cr}. As the temperature inversions in MY30 are not as strong as in MY29, and the tropical tide amplitudes are weaker.

\subsection{Tide structure}

\subsubsection{MY30 in-track data}

\begin{figure}[t]
  \begin{center}
  \includegraphics[width=1.0\linewidth]{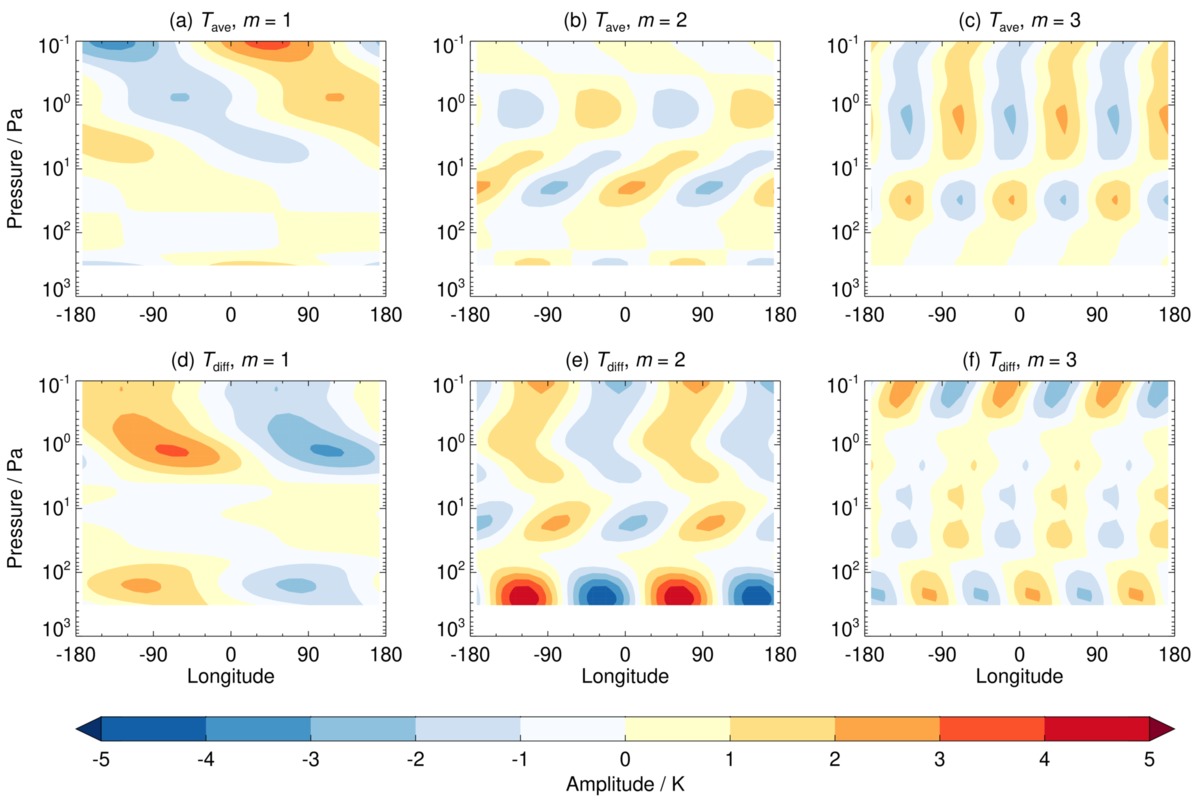}
  \end{center}
  \caption{Amplitude and phase of the \textit{m} = 1--3 components of the (a--c) $T_\mathrm{ave}$, and (d--f) $T_\mathrm{diff}$ fields. Data are from MY30 in-track measurements at a local time of 3:30 am/pm, averaged between 0--10\degree{}N and \lsubs{} = 180--200\degree{}.}
  \label{fig:wave_my30_in}
\end{figure}

To further investigate the differences between the tides in MY29 and MY30, and the changes to the tides with local time in MY30, we look at the longitudinal structure of the tides, in terms of their amplitudes and phases. We begin with the tide structure from the MY30 in-track measurements, as this is similar to all years other than MY29, and so is representative of the general behavior. Figure \ref{fig:wave_my30_in} shows the \textit{m} = 1--3 components of the amplitudes and phases of the $T_\mathrm{ave}$ and $T_\mathrm{diff}$ fields, averaged between 0--10\degree{}N and \lsubs{} = 180--200\degree{}. Corresponding nighttime temperature and water ice distributions are shown in Figure \ref{fig:ice_temp_2my_night}b. Stationary waves and semidiurnal tides are the most likely contributers to the $T_\mathrm{ave}$ field, while diurnal tides are the most likely contributers to the $T_\mathrm{diff}$ field. Higher frequency tides (e.g.\ terdiurnal and quadiurnal) can also potentially contribute, though these generally have smaller amplitudes and a less significant impact on the observed tidal structure \citep{Lee2009, MouddenForbes2014, Guzewich2016}.

In terms of the $T_\mathrm{diff}$ field, the structure of the \textit{m} = 1 component (Figure \ref{fig:wave_my30_in}d) shows evidence for the zonally-symmetric diurnal tide, D0, caused by the higher altitude topography of the Tharsis region. The weak amplitudes and phase shift between $\sim$4--50 Pa are due to the effects of the diurnal tide. This can be seen in Figure S1d, where the \textit{m} = 0 component of $T_\mathrm{diff}$ changes sign, signifying warmer temperatures at 3:30 am compared to 3:30 pm. Outside of the tropics, where the diurnal tide behavior changes and the amplitude of the \textit{s} = 1 component of topography increases, the D0 tide becomes stronger and extends up to $\sim$10 Pa. There is also evidence of a westward-tilted tide, which is most likely DW2. This becomes more prominent at latitudes polewards of $\sim$30\degree{}N (not shown), in tandem with the increasing amplitude of the \textit{s} = 1 component of topography. 

The \textit{m} = 2 component of $T_\mathrm{diff}$ (Figure \ref{fig:wave_my30_in}e) shows evidence of a tide with a long vertical wavelength that is tilted slightly eastward with height, which is suggestive of the DE1 tide forced by the interaction of the diurnal tide with the $s=2$ component of topography. (This tide is most clear in MY33; see Figure S4e.) The DE1 tide has been studied extensively from the surface to the upper atmosphere, with the earliest observations revealing its presence during the 1971 global dust storm \citep{Conrath1976}, and an enhancement in the surface pressure signal in the Viking lander data during the 1977 global dust storms \citep{Leovy1981, Zurek1981, Zurek1988, WilsonHamilton1996, Bridger1998}. Analysis of orbiter data and computer simulations have also revealed the presence of DE1 in the lower and middle atmosphere \citep{WilsonHamilton1996, Wilson2000, Banfield2000, Banfield2003, Hinson2008, Guzewich2012, Guzewich2014} as well as its dominance high in the atmosphere \citep{Wilson2002, Wang2006, Withers2011, Lo2015}.

The \textit{m} = 3 component of $T_\mathrm{diff}$ (Figure \ref{fig:wave_my30_in}f) shows evidence for an eastward propagating tide with a long vertical wavelength, which is most likely DE2. There have been numerous studies of DE2, from the lower atmosphere up to aerobraking altitudes where it is found to play a prominent role in the density structure \citep{Wilson2000, Wilson2002, Banfield2003, Wang2006, Withers2011, Guzewich2012, MouddenForbes2014, Lo2015, Holstein2016}. Superimposed on the DE2 tide's structure are regions of increasing and decreasing amplitudes, which may be the result of the constructive and destructive interference between DE2 and DW4, which is the most likely westward propagating tide. (This DE2 and DW4 interaction is also clear in MY33; see Figure S4.) Analysis of the phase tilts with height in Figure \ref{fig:wave_my30_in}f, as well as in other years, suggests vertical wavelengths of $\sim$3\textit{H} for the westward-traveling wave and $\sim$9\textit{H} for the eastward-traveling wave (where \textit{H} is the atmospheric scale height, with each decade of pressure $\sim$2.3\textit{H}), which are in agreement with the expected wavelengths for the DE2 and DW4 tides \citep{Wilson2000, HinsonWilson2004, MouddenForbes2014, Guzewich2012}. See \citet{Forbes2020} for typical vertical wavelengths of tides based on classical theory. The phase structure near the surface is in agreement with the phase of the $s=3$ component of topography.

The two eastward-propagating tides, DE1 and DE2, play a large role in shaping the nighttime cloud distribution seen in Figure \ref{fig:ice_temp_2my_night}b. At 3:35 am there is a region of clouds centered at $\sim$90\degree{}W, which extends from the surface to $\sim$1 Pa and is tilted slightly eastward with height. There is a similar region in the eastern hemisphere, though cloud opacities are slightly lower here. These cloud regions occur at longitudes where the DE1 and DE2 tides constructively interfere to decrease nighttime temperatures. Elsewhere, temperatures remain too warm, resulting in the cloud-free regions. Where clouds form, local radiative cooling will lead to further decreases in temperature, and possibly further cloud formation, resulting in a positive feedback. This increased nighttime cooling due to clouds is partly responsible for the phase shift between 1--50 Pa in Figure \ref{fig:wave_my30_in}e, which does not correlate with the expected vertical propagation of the surface-forced DE1 tide. Another factor contributing to the phase shift is the anti-symmetric component of the tide (not shown) which tilts eastward with height and has a vertical wavelength of $\sim$6.5\textit{H}, corresponding to the first anti-symmetric component of DE1 \citep{Forbes2020}. The anti-symmetric component is weaker than the symmetric component, but its phase is shifted $\sim$60\degree{} to the east compared to the symmetric component.

The nighttime clouds located at $\sim$90\degree{}W have larger opacities than those at $\sim$90\degree{}E. One reason for this is likely due to the global atmospheric circulation. Stationary waves play a large role in the transport of water away from the subliming north polar ice cap, and the stationary waves over the Tharsis region around southern hemisphere spring equinox tend to transport humid air southwards, while those in the Arabia Terra/Isidis region tend to transport drier air northwards \citep{Steele2014}. As such, the Tharsis region generally has larger water vapor abundances \citep{Smith2002b, Smith2004, Maltagliati2011, Wolkenberg2011, Trokhimovskiy2015}, and hence clouds are likely to be more abundant. Another factor may be the effects of the D0 tide, which results in cooler nighttime temperatures due to the higher topography of the Tharsis region \citep{Banfield2003, Guzewich2012}.

In terms of the $T_\mathrm{ave}$ field, the phase structure of the \textit{m} = 1 component suggests the presence of a tide with a westward phase tilt with height (Figure \ref{fig:wave_my30_in}a), with a vertical wavelength of $\sim$7\textit{H}, possibly SW3. The \textit{m} = 2 component of $T_\mathrm{ave}$ (Figure \ref{fig:wave_my30_in}b) is less organized, and the changing amplitudes and phases with height suggest interference between eastward and westward propagating tides. Analysis of the phase tilts with height in other years, and at other latitudes in the tropics, suggests the westward- and eastward-propagating tides have vertical wavelengths of $\sim$2.5\textit{H} and $\sim$4\textit{H} respectively. It is not possible to identify the exact tides present, but the most likely are S0 and SW4. \citet{Lo2015} previously found evidence for S0 in upper-atmosphere MAVEN observations in mid latitudes. The larger amplitudes between 10--30 Pa may be the result of the radiative cooling associated with the nighttime clouds seen in Figure \ref{fig:ice_temp_2my_night}b. The \textit{m} = 3 component of $T_\mathrm{ave}$ (Figure \ref{fig:wave_my30_in}c) shows a barotropic structure above $\sim$10 Pa. This is likely the SE1 tide, which has previously been observed in both the middle atmosphere \citep{Guzewich2012} and upper atmosphere \citep{Wilson2002, Wang2006, Lo2015}.

\subsubsection{MY30 cross-track data}

\begin{figure}[t]
  \begin{center}
  \includegraphics[width=1.0\linewidth]{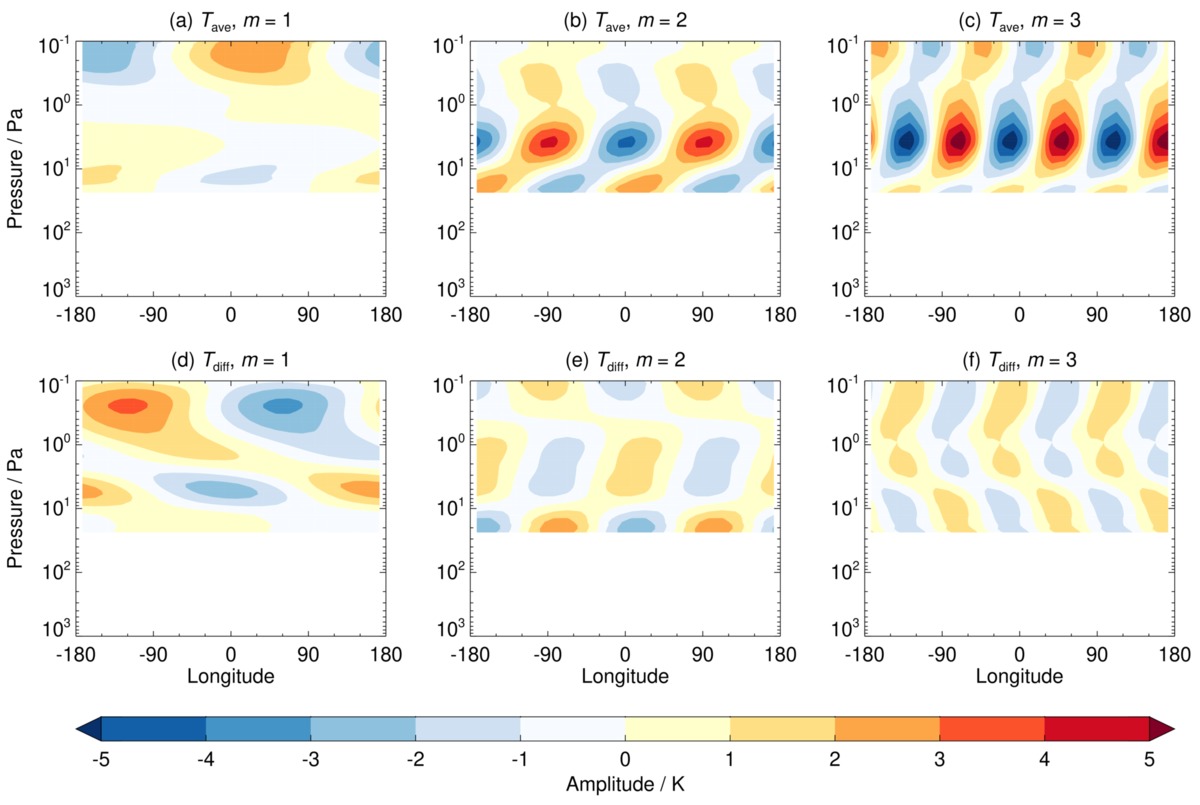}
  \end{center}
  \caption{As Figure \ref{fig:wave_my30_in}, but for MY30 cross-track data at a local time of 5:00 am/pm.}
  \label{fig:wave_my30_cr}
\end{figure}

Figure \ref{fig:wave_my30_cr} shows tide amplitudes and phases in the same format as Figure \ref{fig:wave_my30_in}, but for MY30 cross-track measurements at a local time of $\sim$5:00 am/pm. As noted earlier, the amplitudes of the \textit{m} = 2 and \textit{m} = 3 components of $T_\mathrm{ave}$ (Figure \ref{fig:wave_my30_cr}b,c), and the \textit{m} = 1 component of $T_\mathrm{diff}$ (Figure \ref{fig:wave_my30_cr}d) increase when moving from 03:35 am to 05:00 am, and resemble those in MY29. The corresponding nighttime temperature and water ice distributions are shown in Figure \ref{fig:ice_temp_2my_night}c.

At altitudes below 10 Pa, the \textit{m} = 2 component of $T_\mathrm{ave}$ shows a slight amplification of the tide structure that was present in the in-track data (compare Figures \ref{fig:wave_my30_in}b and \ref{fig:wave_my30_cr}b). This may be from sampling the same tide at a different phase (due to the $\sim$1.5 hour time difference), but the increased amplitudes also correlate with the regions of increased nighttime cloud abundance in Figure \ref{fig:ice_temp_2my_night}c. Thus, there may also be a contribution from local radiative cooling. Between $\sim$1--10 Pa, there is a region of large amplitudes that appear in the cross-track data, but are absent in the in-track data (peak amplitudes are $\sim$5 K at $\sim$4 Pa). This region shows a $\sim$90\degree{} phase shift compared to the tide structure below. Comparing with Figure \ref{fig:ice_temp_2my_night}c, this region of increased amplitudes corresponds to the appearance of nighttime temperature inversions below $\sim$5 Pa.

\begin{figure}[t]
  \begin{center}
  \includegraphics[width=1.0\linewidth]{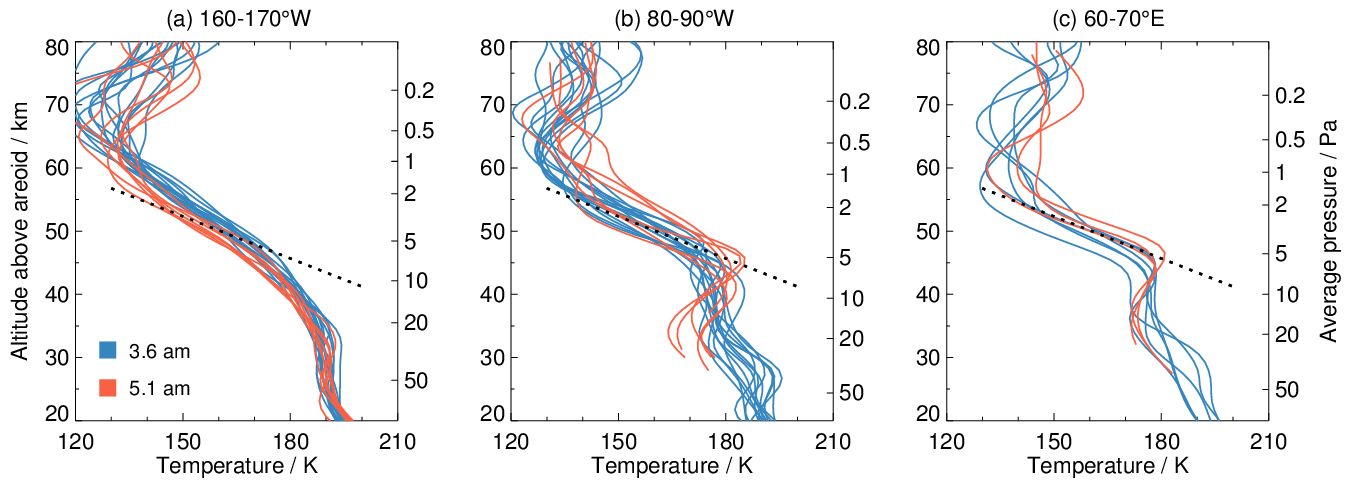}
  \end{center}
  \caption{MY30 temperature profiles at 03:35 am (blue) and 05:00 am (red), from three different longitude regions. Data are averaged between 0--10\degree{}N and \lsubs{} = 185--190\degree{}. The dashed black lines show the adiabatic temperature gradient $g/c_p = 4.5$ K km$^{-1}$.}
  \label{fig:profiles_my30}
\end{figure}

The vertical structure of the temperature inversions is shown in Figure \ref{fig:profiles_my30}, where temperature profiles from three different longitude regions between \lsubs{} = 185--190\degree{} are plotted. (Figure S5 shows the locations of all detected tropical temperature inversions.) Figure \ref{fig:profiles_my30}a shows profiles from a region where there are no large temperature inversions below $\sim$1 Pa, and there is little change in the temperature structure between 03:35 am and 05:00 am. In the 80--90\degree{}W and 60--70\degree{}E regions (Figure \ref{fig:profiles_my30}b,c), temperatures below $\sim$10 Pa ($\sim$40 km) have cooled between 03:35 am and 05:00 am, likely due to cloud radiative effects \citep{Hinson2014}, with greater cooling at 80--90\degree{}W where clouds are thicker. Above the tops of the inversions (above $\sim$5 Pa), the temperatures can be seen to closely follow the adiabatic lapse rate, whereas this is not the case at 160--170\degree{}W, where no inversions occur. The local cooling by clouds, with associated warming above, suggests the increased temperatures are the result of adiabatic warming due to downwelling, with stronger downwelling and increased warming in regions with thicker clouds. This was confirmed via analysis of Ensemble Mars Atmosphere Reanalysis System data (see Appendix \ref{sec:appendix}). This area of adiabatic warming above the cooling cloud region is the reason for the 90\degree{} phase shift at $\sim$10 Pa seen in Figures \ref{fig:wave_my30_cr}b and \ref{fig:wave_my30_cr}e.

In terms of the \textit{m} = 3 component of $T_\mathrm{ave}$, the phases are essentially the same in the in-track and cross-track measurements (compare Figures \ref{fig:wave_my30_in}c and \ref{fig:wave_my30_cr}c), but the peak amplitude has increased from $\sim$3 K to $\sim$6 K between 3:35 am and 05:00 am. This suggests the presence of the same tide as before, SE1, with the larger amplitude the result of sampling the tide at a different phase. The \textit{m} = 1 component of $T_\mathrm{diff}$ (Figure \ref{fig:wave_my30_cr}d) shows increased amplitudes between $\sim$3--8 Pa that appear associated with a westward-tilted tide, with a vertical wavelength of $\sim$3.5\textit{H}. This is in agreement with the expected structure of the DW2 tide \citep{Guzewich2012}. The increased amplitudes correspond to a region of increasing cloud abundance, so there is again likely a contribution from cloud radiative effects. The remaining fields (the \textit{m} = 1 component of $T_\mathrm{ave}$ and the \textit{m} = 3 component of $T_\mathrm{diff}$) do show changes to the tide structure between the in-track and cross-track data, but no large amplitude tides appear, and as such we do not discuss these components further.

\subsubsection{MY29 in-track data}

Comparing Figures \ref{fig:ice_temp_2my_night}b--d, the MY29 nighttime temperature and water ice distributions at 03:45 am more closely resemble the MY30 05:00 am data, as opposed to the MY30 03:35 data, despite the latter being closer in local time. As the temperature and ice distributions are largely influenced by diurnal tides, this suggests the early dust activity in MY29 has shifted the phase of the diurnal tide to an earlier local time. Such a shift, as a response to increased dustiness during global dust storms, has been noted previously in modeling and observations \citep[e.g.][]{WilsonRichardson2000, Guzewich2016}. However, the early dust activity in MY29 did not produce large dust optical depths like those experienced in global dust storms.

\begin{figure}[t]
  \begin{center}
  \includegraphics[width=1.0\linewidth]{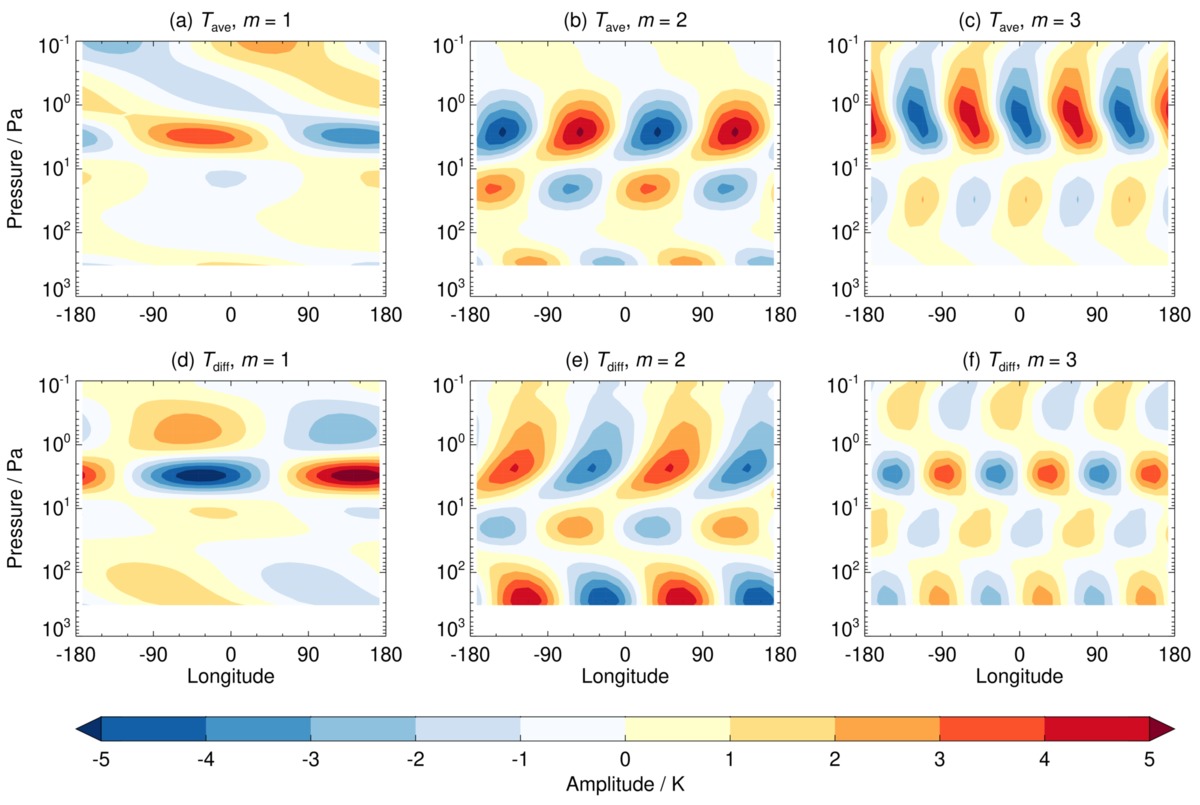}
  \end{center}
  \caption{As Figure \ref{fig:wave_my30_in}, but for MY29 in-track data at a local time of 3:45 am/pm.}
  \label{fig:wave_my29_in}
\end{figure}

The amplitudes and phases of the tides in MY29 between \lsubs{} = 180--200\degree{} are shown in Figure \ref{fig:wave_my29_in}. There is a general eastwards shift in the \textit{m} = 2 and \textit{m} = 3 tide phases compared to MY30 (Figure \ref{fig:wave_my30_in}), which results in the clouds in Figure \ref{fig:ice_temp_2my_night}d being located further east than in MY30 (Figure \ref{fig:ice_temp_2my_night}b,c). This phase shift corresponds to an increase in the dust abundance of the low-lying planitias, which shifts the largest column dust opacities eastward between \lsubs{} = 150--170\degree{}. The dust distribution then remains steady until \lsubs{} = 200\degree{}, and correspondingly the tide phases show little change during this time. There is also an increase in the amplitudes of the westward-propagating tide components. These were most prominent after the onset of the early dust activity, but began to decrease in amplitude between \lsubs{} = 150--170\degree{} as the dust lofted by the storm began decreasing in altitude. Despite this eastward shift, the tide structure in MY29 below $\sim$10 Pa generally resembles that in the MY30 in-track data (Figure \ref{fig:wave_my30_in}).

\begin{figure}[t]
  \begin{center}
  \includegraphics[width=1.0\linewidth]{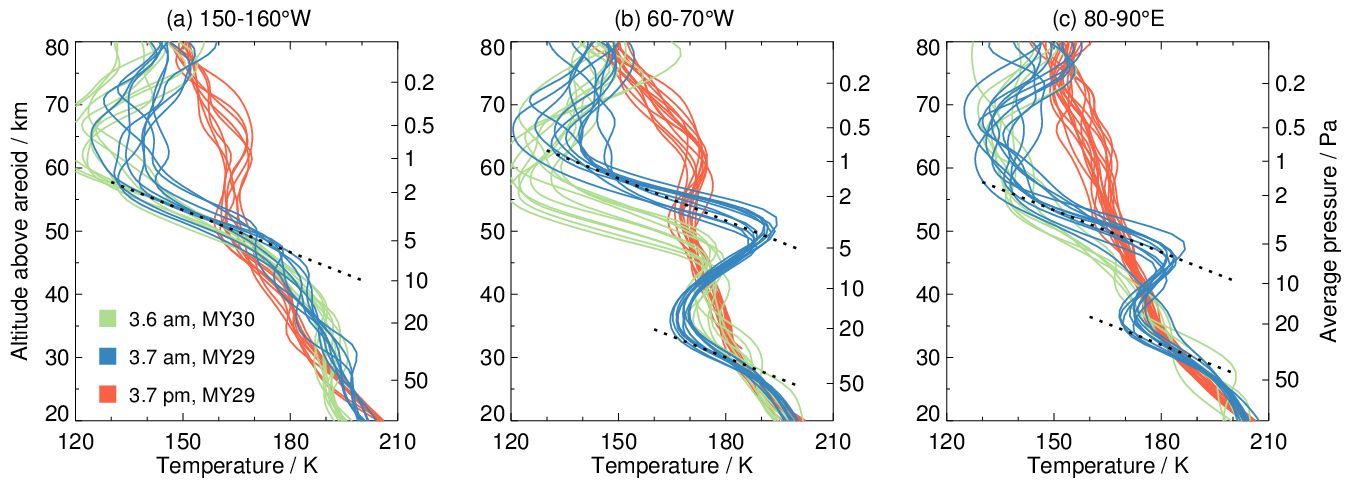}
  \end{center}
  \caption{Temperature profiles from MY30 nighttime (green), MY29 nighttime (blue), and MY29 daytime (red) in-track measurements. Data are shown for three different longitude regions, and are averaged between 0--10\degree{}N and \lsubs{} = 190--195\degree{}. The dashed black lines show the adiabatic temperature gradient.}
  \label{fig:profiles_2my_in}
\end{figure}

As in the MY30 cross-track data, nighttime temperature inversions occur in MY29, but these are much stronger than those in MY30, particularly in the western hemisphere.
Figure \ref{fig:profiles_2my_in} shows individual temperature profiles from three different longitude regions between \lsubs{} = 190--195\degree{}, when the inversions are strongest. (Figure S6 shows the locations of all detected tropical temperature inversions.) Data are shown from 03:35 am in MY30, and 03:45 am/pm in MY29. The daytime temperatures are similar in all three regions, as are the nighttime temperatures above $\sim$1 Pa. In Figure \ref{fig:profiles_2my_in}a, where no large temperature inversions occur below $\sim$1 Pa, the nighttime temperatures in MY29 and MY30 are generally similar. In the 60--70\degree{}W region, which has the strongest inversions (Figure \ref{fig:profiles_2my_in}b), the nighttime temperatures increase by $\sim$25 K while ascending from 35--50 km. The coolest temperatures at the base of the inversions at 35 km are $\sim$20 K cooler compared to the 150--160\degree{}W region, and the warmest temperatures at 50 km are $\sim$25 K warmer. The 80--90\degree{}E region has weaker temperature inversions (Figure \ref{fig:profiles_2my_in}b), with peak temperatures occurring at $\sim$45 km, compared to $\sim$50 km at 60--70\degree{}W.

As in the MY30 cross-track data at 05:00 am (Figure \ref{fig:profiles_my30}), temperatures closely follow the adiabatic lapse rate above the tops of the inversions, suggesting adiabatic warming due to downwelling. However, the temperature inversions are much stronger in MY29 than in MY30, which suggests the diurnal and semidiurnal tides have been strengthened in the regions with inversions. The diurnal and semidiurnal tides have been observed to strengthen during global dust storms \citep{LeovyZurek1979, Zurek1980, Bridger1998, Guzewich2016, Guzewich2019, Viudez2019}, while more recent studies have revealed the semidiurnal tide is prominent all year, and is influenced by the distribution of water ice clouds as well as dust \citep{Kleinbohl2013, Guzewich2016, Haberle2019}. Due to the increased coverage to lower altitudes compared to MY30 cross-track measurements, the temperatures below the bases of the inversions in MY29 can also be seen to follow the adiabatic temperature gradient (Figure \ref{fig:profiles_2my_in}b,c), showing that this is a region of convective instability \citep{Hinson2014}. This instability may cause water vapor (as well as dust and any other constituents present) from lower in the atmosphere to be transported upwards, increasing the cloud opacity and further strengthening the local cooling, leading to a positive feedback.

The nighttime temperature inversions in MY29 begin to appear at \lsubs{} $\approx$ 145\degree{}, and they strengthen and expand longitudinally until \lsubs{} $\approx$ 200\degree{} (see Figure S6), in tandem with a general warming of the atmosphere. The increasing temperatures and longitudinal spreading of the regions with temperature inversions suggest that dust is playing a role in their formation. The dust abundance rapidly increases after the onset of the early dust activity, but then the height of the dust top falls between \lsubs{} $\approx$ 150--170\degree{}, before slowly rising again after \lsubs{} $\approx$ 170\degree{}. This more slow rising is likely due to upwelling in the Hadley cells, as the twin Hadley cell structure moves equatorward from the onset of the dust activity at \lsubs{} $\approx$ 143\degree{} to \lsubs{} $\approx$ 200\degree{} \citep{Steele2014}. The increase in dust height and opacity will lead to a strengthening of the diurnal and semidiurnal tides, which can in turn impact the non-migrating thermal tides. The inversions begin to decline between \lsubs{} $\approx$ 200--210\degree{}, during which time the circulation changes from a twin Hadley cell structure with upwelling in the tropics to an equator-crossing Hadley cell.

As seen in Figure \ref{fig:ampl_my29_in}, apart from the \textit{m} = 3 component of $T_\mathrm{ave}$, the large amplitude tides in the tropics in MY29 are focused in the 2--5 Pa region, and biased to the northern tropics, suggesting the tides are linked to a similar forcing mechanism. Also, from the analysis of MY30 cross-track data at local times of 05:00 am/pm, when the ice cloud distribution is a good match for that in MY29, only the \textit{m} = 2 and \textit{m} = 3 components of $T_\mathrm{diff}$ show a large amplitude increase (Figure \ref{fig:ampl_my30_cr}). This suggests that the dust distribution in MY29 is playing an important role in strengthening the tides. Indeed, water ice clouds have been shown to influence the semidiurnal tide \citep{Kleinbohl2013}, which plays a key role in the structure of the $T_\mathrm{ave}$ field. This is likely why the tides in the $T_\mathrm{ave}$ field show an increase with local time in MY30, as the ice cloud distribution changes over the $\sim$1.5 hour period, whereas the dust does not. However, the $T_\mathrm{diff}$ field is affected mainly by the diurnal tide, which is known to be sensitive to local dust loading \citep{LeovyZurek1979, Bridger1998, Guzewich2016}. This is likely why the $T_\mathrm{diff}$ field shows stronger tides in MY29 compared to MY30, as the atmospheric dust content was larger. 

The \textit{m} = 1 components of both $T_\mathrm{ave}$ and $T_\mathrm{diff}$ have peak amplitudes at $\sim$30--40\degree{}W. This corresponds to a region of low-lying topography to the east of the Tharsis plateau and south of Acidalia Planita, but does not correlate to a region of high or low amplitudes in the \textit{s} = 1 component of topography, ruling out surface forcing of the tides. However, this region is a favored location for equator-crossing dust storms \citep{Cantor2007, WangRichardson2015, Shirley2020}, and the early dust activity in MY29 occurred in this region. Due to the low-lying topography, the column dust abundances here are larger, resulting in a stronger forcing of the \textit{m} = 1 components of the tide. The phase change at $\sim$3 Pa is related to the nighttime warming associated with the stronger downwelling. 

The tide structures in the \textit{m} = 2 components of $T_\mathrm{ave}$ are similar to those in the MY30 cross-track data (Figure \ref{fig:wave_my30_cr}b,e). The eastward tilt in the $T_\mathrm{diff}$ field is from an anti-symmetric tidal mode (not shown), with a wavelength of $\sim$6\textit{H}. This is comparable to the first anti-symmetric component of DE1 \citep{Forbes2020}. This anti-symmetric component exists with a similar amplitude in all years, but in MY29 it constructively interferes with the symmetric component, causing an amplification of the tide above 10 Pa. The large-amplitude tide in the \textit{m} = 3 component of $T_\mathrm{ave}$ shows a westward phase tilt compared to MY30. The lower atmosphere shows evidence for a westward-propagating wave with a vertical wavelength of $\sim$3\textit{H}, possibly SW5, which may be interacting with SE1 and causing the observed tilt. The region of large amplitudes in the \textit{m} = 3 component of $T_\mathrm{diff}$, centered at $\sim$3 Pa, appears to result from an amplification of the westward-propagating DW4 tide, a weakening of the eastward-propagating DE2 tide, or both. 


\section{Conclusions}\label{sec:conclusions}

The results of the Fourier analysis of MY29 in-track temperatures, at a local time of $\sim$03:45 am/pm around southern hemisphere spring equinox, reveal large amplitude non-migrating tides in the tropics that are not present in any other Mars year observed by MCS. MY29 experienced early dust activity, beginning at \lsubs{} $\approx$ 143\degree{}, and the observed tides began to strengthen after this dust activity, reaching their largest amplitudes at \lsubs{} $\approx$ 190\degree{}. Comparisons between daytime and nighttime temperatures show that it is the nighttime temperature distribution that is most perturbed in MY29, with strong temperature inversions located at $\sim$35--55 km in the 15--90\degree{}W and 75--110\degree{}E regions, though these regions expand eastward with time between \lsubs{} = 185--200\degree{}. Due to this observed longitudinal spreading with \lsubs{}, the temperature inversions in MY29 appear to be related not only to the topography, but also to the aerosol distribution.

While the temperature inversions and tides continue to strengthen until \lsubs{} $\approx$ 190\degree{} in MY29, the early dust storm activity ceased at \lsubs{} $\approx$ 157\degree{} \citep{WangRichardson2015}. The continued strengthening of the tropical tides, and the longitudinal expansion of the regions displaying temperature inversions, suggests that dust continued to be transported equatorward and vertically. This is plausible, as the twin Hadley cell circulation has its upwelling branch over the tropics during southern hemisphere spring equinox. Transport by the Hadley cell circulation, aided by thermal tides, has been suggested as a mechanism for dust entering the northern jet and encircling Mars during the MY34 global dust storm, which also occurred during southern hemisphere spring equinox \citep{Gillespie2020}. 

When analyzing MCS cross-track data in other years, taken at later local times of $\sim$05:00 am/pm, some stronger tides do appear, notably in the \textit{m} = 2 and \textit{m} = 3 components of $T_\mathrm{ave}$, where the tide structures resemble those in MY29. The $T_\mathrm{ave}$ field can have contributions from stationary waves, but the phases and inferred vertical wavelengths seem to suggest that it is the semidiurnal non-migrating tides that are amplified at later local times. In the case of the \textit{m} = 3 component of $T_\mathrm{ave}$, the amplification is likely due to sampling the eastward-propagating SE1 tide at a different local time, with aerosols having little impact. For the \textit{m} = 2 component of $T_\mathrm{ave}$, the tide amplification appears to be related to the water ice cloud distribution. The radiative cooling associated with regions of higher-opacity nighttime clouds causes stronger downwelling of the air aloft, and increased adiabatic warming. The local cooling at cloud level and warming above results in the appearance of nighttime temperature inversions.

The nighttime cloud structure, which results in the strengthening of the semidiurnal tide, is itself strongly influenced by the diurnal non-migrating tides, particularly the combination of the eastward-propagating DE1 and DE2 tides. Where these tides constructively interfere to decrease nighttime temperatures, clouds preferentially form, leading to increased radiative cooling. Due to this cooling, the bases of the temperature inversions are regions of convective instability, which could result in water vapor (as well as dust and any other constituents present) from lower in the atmosphere being transported upwards, increasing the cloud opacity and further strengthening the local cooling, leading to a positive feedback.

Comparing different Mars years and different local times, it appears as though the radiative influence of water ice clouds is impacting the semidiurnal non-migrating tides in MY29 (and other years), while the large-amplitude diurnal non-migrating tides in MY29 are influenced by the early dust activity. The dust in MY29 was concentrated mostly in the northern tropics, which explains why the large tide amplitudes are biased towards this region. The fact that nighttime temperature inversions occur at earlier local times in MY29 (03:45 am) compared to other years (05:00 am) suggests there was a shift in the phase of the diurnal tide to an earlier local time. Such a shift, as a response to increased dustiness during global dust storms, has been noted previously in modeling and observations \citep[e.g.][]{WilsonRichardson2000, Guzewich2016}. 

While the early dust activity in MY29 did not produce large dust optical depths like those experienced in global dust storms, the dust activity occurred at a time when there was upwelling over the tropics, associated with the Hadley circulation, enabling the dust to be transported to higher altitudes where it has a larger radiative influence. The warmer tropical temperatures associated with the dust led to a strengthening of the Hadley circulation (as determined from the increased temperatures in the polar warmings above both polar vortices), resulting in a positive feedback. Combined with this increased vertical transport of dust, southern hemisphere spring equinox is when the water vapor column abundances in the tropics reach their largest value over the year \citep{Smith2004, Heavens2011c, Wolkenberg2011, Steele2014, Trokhimovskiy2015}. As such, water vapor is also available for transport, which can result in increased cloud formation, which in turn can impact the tidal structure through local radiative heating and cooling. The tide amplitudes and temperature inversion strengths in MY29 begin to decrease after \lsubs{} $\approx$ 190\degree{}, in combination with a decrease in the dustiness of the tropics, a change in the overturning circulation to an equator-crossing Hadley cell, and a decrease in the water vapor abundance.


\appendix
\renewcommand\thefigure{\thesection\arabic{figure}}
\setcounter{figure}{0}    
\section{Temperature inversions in EMARS data}\label{sec:appendix}

In order to test the hypothesis that downwelling associated with the diurnal tide can cause the temperature inversions observed in the MCS measurements, we use Ensemble Mars Atmosphere Reanalysis System (EMARS) version 1.0 data \citep{Greybush2019} for MY29. This dataset results from the assimilation of MCS in-track temperature profiles using a Local Ensemble Transform Kalman Filter \citep[LETKF; see][]{Hoffman2010, Greybush2012, Zhao2015}, and is freely available at \url{https://www.datacommons.psu.edu/commonswizard/MetadataDisplay.aspx?Dataset=6171}. Dust and water ice tracers are advected by the model winds, and are radiatively active. However, for dust there is a source/sink term in the boundary layer that relaxes the model column opacities towards the dust maps of \citet{Montabone2015}. We use the `background' files, which result from 1-hour forecasts initiated from the LETKF analyses, as the analysis files only contain state variables and hence provide no information on the aerosol distributions. We temporally interpolate the hourly data in each grid box to obtain values at the same local time at each longitude.

\begin{figure}[t]
  \begin{center}
  \includegraphics[width=0.95\linewidth]{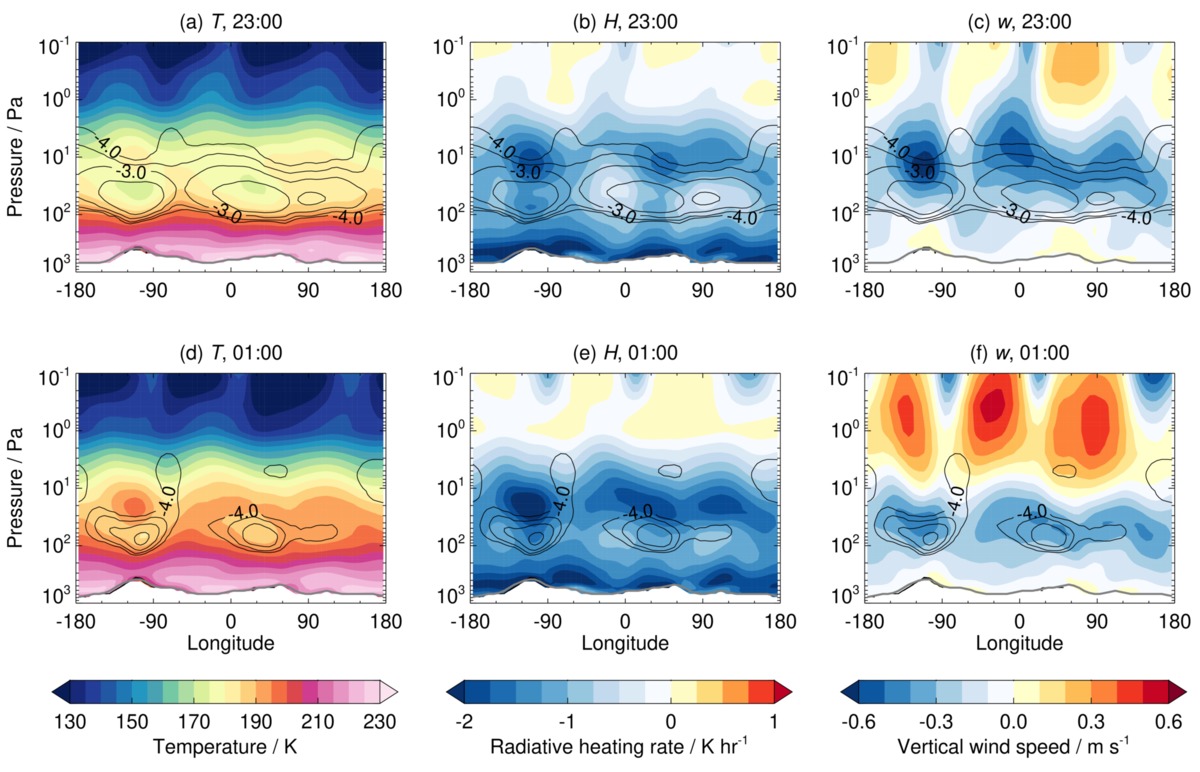}
  \end{center}
  \caption{EMARS MY29 simulation results of (a,d) temperature, (b,e) radiative heating rate, and (c,f) vertical wind speed at 11:00 pm and 1:00 am. Data are averaged between 0--10\degree{}N over \lsubs{} = 180--200\degree{}. Black contours show water ice opacity per km in log units, in steps of 0.5.}
  \label{fig:loct_emars}
\end{figure}

Nighttime temperature inversions exist in the EMARS dataset in the \lsubs{} = 180--200\degree{} period, but they occur earlier in the night and lower in the atmosphere compared to the MCS data. The inversions begin forming at around $\sim$9:00 pm, strengthen the most between 11:00 pm and 1:00 am, and disappear just after 3:00 am. The maximum temperature in the inversion over the Tharsis region increases by $\sim$20 K between 11:00 pm and 1:00 am, while the temperature minima below the inversion increases by $\sim$15 K. Figure \ref{fig:loct_emars} shows temperatures, radiative heating rates and vertical winds at 11:00 pm and 1:00 am. The heating rate is due to all gases and aerosols combined, as EMARS does not split the contributions from different species. It is clear that heating rates are negative everywhere below $\sim$1 Pa, resulting in a cooling of the atmosphere. Thus, some other process must be responsible for the increased temperatures in the inversions. At this time the vertical wind speed, \textit{w}, has negative values around the altitude of the inversions. This indicates downwelling in the atmosphere, and is strongest in regions with larger ice opacities. (Vertical winds were calculated from $\omega = \mathrm{d}p/\mathrm{d}t$, as output by EMARS, by assuming hydrostatic equilibrium.) The overall downward progression with time is associated with the diurnal tide, which is amplified by the radiative influence of clouds \citep{WilsonRichardson2000, Wilson2014}. Due to compression of the air as it moves downwards, its temperature increases due to adiabatic warming \citep{HinsonWilson2004, Wilson2014}. 

\begin{figure}[t]
  \begin{center}
  \includegraphics[width=1.0\linewidth]{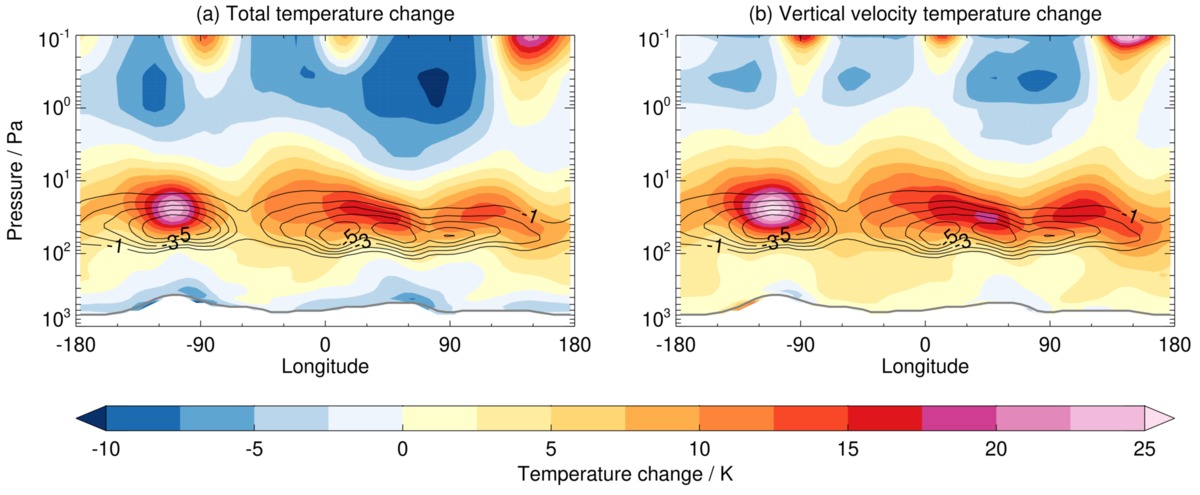}
  \end{center}
  \caption{(a) Temperature change between 23:00 and 01:00 from EMARS MY29 simulation results, averaged between 0--10\degree{}N over \lsubs{} = 180--200\degree{}. (b) Temperature change obtained by taking the 23:00 temperature and applying only the adiabatic dynamical heating from the vertical velocity. Black contours show the change in water ice opacity, in units of 10$^{-3}$ km$^{-1}$}
  \label{fig:emars_dyn_heat}
\end{figure}

To investigate the potential warming caused by downwelling, we take the 11:00 pm temperatures, and using only the adiabatic temperature changes caused by the vertical wind field (ignoring radiative heating or cooling) we calculate new temperatures in 5 minute intervals until 01:00 am. We do this at each grid point by using the vertical wind to determine the altitude of an air parcel 5 minutes earlier. We then interpolate the temperature profile to find the temperature at this location, $T_\mathrm{init}$. Over the time interval $\Delta t$ we then calculate the temperature change as $\Delta T = -w \Delta t g / c_p$, assuming a dry adiabatic lapse rate of $g/c_p = 4.5$ K km$^{-1}$. The new temperature at the grid point is then $T_\mathrm{new} = T_\mathrm{init} + \Delta T$. The results, after two hours of iteration, are shown in Figure \ref{fig:emars_dyn_heat}. As can be seen, most of the temperature changes that occur away from the surface between 11:00 pm and 1:00 am in the EMARS data can be attributed to the adiabatic heating or cooling of the air in response to the vertical winds. The temperature changes are slightly too large in the Tharsis region as radiative cooling is not taken into account. Between 3:00--5:00 am the vertical wind around the cloud layer becomes positive, and adiabatic heating ceases. Thus, it is likely that the large temperatures seen in the inversions in the MCS data are the result of increased downwelling due to stronger tides forced by cloud radiative effects.


\section*{Acknowledgements}

The authors are grateful to Jeffrey Forbes for fruitful discussions about thermal tides, and to Sylvain Piqueux and Nicholas Heavens for helpful comments on the manuscript. The data used here are freely available from NASA's Planetary Data System at \url{https://atmos.nmsu.edu/data\_and\_services/atmospheres\_data/MARS/mcs.html}. This work was performed at the Jet Propulsion Laboratory, California Institute of Technology, under a contract with NASA. Copyright 2020, California Institute of Technology. Government sponsorship acknowledged.


\singlespacing
\bibliography{aerosol_tides}

\begin{thebibliography}{}

\bibitem[{Banfield} et~al., 2000]{Banfield2000}
{Banfield}, D., {Conrath}, B., {Pearl}, J.~C., {Smith}, M.~D., and
  {Christensen}, P. (2000).
\newblock {Thermal tides and stationary waves on Mars as revealed by Mars
  Global Surveyor thermal emission spectrometer}.
\newblock {\em J. Geophys. Res.}, 105(E4):9521--9538.

\bibitem[{Banfield} et~al., 2003]{Banfield2003}
{Banfield}, D., {Conrath}, B.~J., {Smith}, M.~D., {Christensen}, P.~R., and
  {Wilson}, R.~J. (2003).
\newblock {Forced waves in the martian atmosphere from MGS TES nadir data}.
\newblock {\em Icarus}, 161(2):319--345.

\bibitem[{Bougher} et~al., 2004]{Bougher2004}
{Bougher}, S.~W., {Engel}, S., {Hinson}, D.~P., and {Murphy}, J.~R. (2004).
\newblock {MGS Radio Science electron density profiles: Interannual variability
  and implications for the Martian neutral atmosphere}.
\newblock {\em J. Geophys. Res. (Planets)}, 109(E3):E03010.

\bibitem[{Bridger} and {Murphy}, 1998]{Bridger1998}
{Bridger}, A. F.~C. and {Murphy}, J.~R. (1998).
\newblock {Mars' surface pressure tides and their behavior during global dust
  storms}.
\newblock {\em J. Geophys. Res.}, 103(E4):8587--8602.

\bibitem[{Cahoy} et~al., 2006]{Cahoy2006}
{Cahoy}, K.~L., {Hinson}, D.~P., and {Tyler}, G.~L. (2006).
\newblock {Radio science measurements of atmospheric refractivity with Mars
  Global Surveyor}.
\newblock {\em J. Geophys. Res. (Planets)}, 111(E5):E05003.

\bibitem[{Cantor}, 2007]{Cantor2007}
{Cantor}, B.~A. (2007).
\newblock {MOC observations of the 2001 Mars planet-encircling dust storm}.
\newblock {\em Icarus}, 186(1):60--96.

\bibitem[{Conrath}, 1976]{Conrath1976}
{Conrath}, B.~J. (1976).
\newblock {Influence of planetary-scale topography on the diurnal thermal tide
  during the 1971 Martian dust storm.}
\newblock {\em J. Atmos. Sci.}, 33:2430--2439.

\bibitem[{England} et~al., 2016]{England2016}
{England}, S.~L., {Liu}, G., {Withers}, P., {Yi{\v{g}}it}, E., {Lo}, D.,
  {Jain}, S., {Schneider}, N.~M., {Deighan}, J., {McClintock}, W.~E.,
  {Mahaffy}, P.~R., {Elrod}, M., {Benna}, M., and {Jakosky}, B.~M. (2016).
\newblock {Simultaneous observations of atmospheric tides from combined in situ
  and remote observations at Mars from the MAVEN spacecraft}.
\newblock {\em J. Geophys. Res. (Planets)}, 121(4):594--607.

\bibitem[{Forbes}, 1995]{Forbes1995}
{Forbes}, J.~M. (1995).
\newblock {Tidal and Planetary Waves}.
\newblock {\em Washington DC American Geophysical Union Geophysical Monograph
  Series}, 87:67.

\bibitem[{Forbes}, 2002]{Forbes2002}
{Forbes}, J.~M. (2002).
\newblock {Wave Coupling in Terrestrial Planetary Atmospheres}.
\newblock {\em Washington DC American Geophysical Union Geophysical Monograph
  Series}, 130:171.

\bibitem[{Forbes} and {Hagan}, 2000]{ForbesHagan2000}
{Forbes}, J.~M. and {Hagan}, M.~E. (2000).
\newblock {Diurnal Kelvin wave in the atmosphere of Mars: Towards an
  understanding of {\textquoteleft}stationary{\textquoteright} density
  structures observed by the MGS accelerometer}.
\newblock {\em Geophys. Res. Lett.}, 27(21):3563--3566.

\bibitem[{Forbes} and {Miyahara}, 2006]{ForbesMiyahara2006}
{Forbes}, J.~M. and {Miyahara}, S. (2006).
\newblock {Solar Semidiurnal Tide in the Dusty Atmosphere of Mars.}
\newblock {\em J. Atmos. Sci.}, 63(7):1798--1817.

\bibitem[{Forbes} et~al., 2020]{Forbes2020}
{Forbes}, J.~M., {Zhang}, X., {Forget}, F., {Millour}, E., and {Kleinb{\"o}hl},
  A. (2020).
\newblock {Solar Tides in the Middle and Upper Atmosphere of Mars}.
\newblock {\em J. Geophy. Res. Space Phys.}, 125(9).

\bibitem[{Gillespie} et~al., 2020]{Gillespie2020}
{Gillespie}, H.~E., {Greybush}, S.~J., and {Wilson}, R.~J. (2020).
\newblock {An Investigation of the Encirclement of Mars by Dust in the 2018
  Global Dust Storm Using EMARS}.
\newblock {\em J. Geophys. Res. (Planets)}, 125(7):e06106.

\bibitem[{Greybush} et~al., 2019]{Greybush2019}
{Greybush}, S.~J., {Kalnay}, E., {Wilson}, R.~J., {Hoffman}, R.~N., {Nehrkorn},
  T., {Leidner}, M., {Eluszkiewicz}, J., {Gillespie}, H.~E., {Wespetal}, M.,
  {Zhao}, Y., {Hoffman}, M.~J., {Dudas}, P., {McConnochie}, T.,
  {Kleinb{\"o}hl}, A., {Kass}, D.~M., {McCleese}, D.~J., and {Miyoshi}, T.
  (2019).
\newblock {The Ensemble Mars Atmosphere Reanalysis System (EMARS) Version 1.0}.
\newblock {\em Geosci. Data J.}, 6(2):137--150.

\bibitem[{Greybush} et~al., 2012]{Greybush2012}
{Greybush}, S.~J., {Wilson}, R.~J., {Hoffman}, R.~N., {Hoffman}, M.~J.,
  {Miyoshi}, T., {Ide}, K., {McConnochie}, T., and {Kalnay}, E. (2012).
\newblock {Ensemble Kalman filter data assimilation of Thermal Emission
  Spectrometer temperature retrievals into a Mars GCM}.
\newblock {\em J. Geophys. Res. (Planets)}, 117(E11):E11008.

\bibitem[{Guzewich} et~al., 2019]{Guzewich2019}
{Guzewich}, S.~D., {Lemmon}, M., {Smith}, C.~L., {Mart{\'\i}nez}, G., {de
  Vicente-Retortillo}, {\'A}., {Newman}, C.~E., {Baker}, M., {Campbell}, C.,
  {Cooper}, B., {G{\'o}mez-Elvira}, J., {Harri}, A.~M., {Hassler}, D.,
  {Martin-Torres}, F.~J., {McConnochie}, T., {Moores}, J.~E.,
  {Kahanp{\"a}{\"a}}, H., {Khayat}, A., {Richardson}, M.~I., {Smith}, M.~D.,
  {Sullivan}, R., {de la Torre Juarez}, M., {Vasavada}, A.~R.,
  {Vi{\'u}dez-Moreiras}, D., {Zeitlin}, C., and {Zorzano Mier}, M.-P. (2019).
\newblock {Mars Science Laboratory Observations of the 2018/Mars Year 34 Global
  Dust Storm}.
\newblock {\em Geophys. Res. Lett.}, 46(1):71--79.

\bibitem[{Guzewich} et~al., 2016]{Guzewich2016}
{Guzewich}, S.~D., {Newman}, C.~E., {de la Torre Ju{\'a}rez}, M., {Wilson},
  R.~J., {Lemmon}, M., {Smith}, M.~D., {Kahanp{\"a}{\"a}}, H., and {Harri},
  A.~M. (2016).
\newblock {Atmospheric tides in Gale Crater, Mars}.
\newblock {\em Icarus}, 268:37--49.

\bibitem[{Guzewich} et~al., 2012]{Guzewich2012}
{Guzewich}, S.~D., {Talaat}, E.~R., and {Waugh}, D.~W. (2012).
\newblock {Observations of planetary waves and nonmigrating tides by the Mars
  Climate Sounder}.
\newblock {\em J. Geophys. Res. (Planets)}, 117(E3):E03010.

\bibitem[{Guzewich} et~al., 2014]{Guzewich2014}
{Guzewich}, S.~D., {Wilson}, R.~J., {McConnochie}, T.~H., {Toigo}, A.~D.,
  {Banfield}, D.~J., and {Smith}, M.~D. (2014).
\newblock {Thermal tides during the 2001 Martian global-scale dust storm}.
\newblock {\em J. Geophys. Res. (Planets)}, 119(3):506--519.

\bibitem[{Haberle} et~al., 2019]{Haberle2019}
{Haberle}, R.~M., {Kahre}, M.~A., {Hollingsworth}, J.~L., {Montmessin}, F.,
  {Wilson}, R.~J., {Urata}, R.~A., {Brecht}, A.~S., {Wolff}, M.~J., {Kling},
  A.~M., and {Schaeffer}, J.~R. (2019).
\newblock {Documentation of the NASA/Ames Legacy Mars Global Climate Model:
  Simulations of the present seasonal water cycle}.
\newblock {\em Icarus}, 333:130--164.

\bibitem[{Heavens} et~al., 2011]{Heavens2011c}
{Heavens}, N.~G., {McCleese}, D.~J., {Richardson}, M.~I., {Kass}, D.~M.,
  {Kleinb{\"o}hl}, A., and {Schofield}, J.~T. (2011).
\newblock {Structure and dynamics of the Martian lower and middle atmosphere as
  observed by the Mars Climate Sounder: 2. Implications of the thermal
  structure and aerosol distributions for the mean meridional circulation}.
\newblock {\em J. Geophys. Res. (Planets)}, 116(E1):E01010.

\bibitem[{Hinson} et~al., 2014]{Hinson2014}
{Hinson}, D.~P., {Asmar}, S.~W., {Kahan}, D.~S., {Akopian}, V., {Haberle},
  R.~M., {Spiga}, A., {Schofield}, J.~T., {Kleinb{\"o}hl}, A., {Abdou}, W.~A.,
  {Lewis}, S.~R., {Paik}, M., and {Maalouf}, S.~G. (2014).
\newblock {Initial results from radio occultation measurements with the Mars
  Reconnaissance Orbiter: A nocturnal mixed layer in the tropics and
  comparisons with polar profiles from the Mars Climate Sounder}.
\newblock {\em Icarus}, 243:91--103.

\bibitem[{Hinson} et~al., 2008]{Hinson2008}
{Hinson}, D.~P., {P{\"a}tzold}, M., {Wilson}, R.~J., {H{\"a}usler}, B.,
  {Tellmann}, S., and {Tyler}, G.~L. (2008).
\newblock {Radio occultation measurements and MGCM simulations of Kelvin waves
  on Mars}.
\newblock {\em Icarus}, 193(1):125--138.

\bibitem[{Hinson} and {Wilson}, 2004]{HinsonWilson2004}
{Hinson}, D.~P. and {Wilson}, R.~J. (2004).
\newblock {Temperature inversions, thermal tides, and water ice clouds in the
  Martian tropics}.
\newblock {\em J. Geophys. Res. (Planets)}, 109(E1):E01002.

\bibitem[{Hoffman} et~al., 2010]{Hoffman2010}
{Hoffman}, M.~J., {Greybush}, S.~J., {John Wilson}, R., {Gyarmati}, G.,
  {Hoffman}, R.~N., {Kalnay}, E., {Ide}, K., {Kostelich}, E.~J., {Miyoshi}, T.,
  and {Szunyogh}, I. (2010).
\newblock {An ensemble Kalman filter data assimilation system for the martian
  atmosphere: Implementation and simulation experiments}.
\newblock {\em Icarus}, 209(2):470--481.

\bibitem[{Holstein-Rathlou} et~al., 2016]{Holstein2016}
{Holstein-Rathlou}, C., {Maue}, A., and {Withers}, P. (2016).
\newblock {Atmospheric studies from the Mars Science Laboratory Entry, Descent
  and Landing atmospheric structure reconstruction}.
\newblock {\em Planet. Space Sci.}, 120:15--23.

\bibitem[{Jiang} et~al., 2019]{Jiang2019}
{Jiang}, F.~Y., {Yelle}, R.~V., {Jain}, S.~K., {Cui}, J., {Montmessin}, F.,
  {Schneider}, N.~M., {Deighan}, J., {Gr{\"o}ller}, H., and {Verdier}, L.
  (2019).
\newblock {Detection of Mesospheric CO$_{2}$ Ice Clouds on Mars in Southern
  Summer}.
\newblock {\em Geophys. Res. Lett.}, 46(14):7962--7971.

\bibitem[{Joshi} et~al., 2000]{Joshi2000}
{Joshi}, M.~M., {Hollingsworth}, J.~L., {Haberle}, R.~M., and {Bridger}, A.
  F.~C. (2000).
\newblock {An interpretation of Martian thermospheric waves based on analysis
  of a general circulation model}.
\newblock {\em Geophys. Res. Lett.}, 27(5):613--616.

\bibitem[{Kass} et~al., 2019]{Kass2019}
{Kass}, D., {Schofield}, J., {Kleinb{\"o}hl}, A., {McCleese}, D., {Heavens},
  N., {Shirley}, J., and {Steele}, L. (2019).
\newblock {Mars Climate Sounder observation of Mars' 2018 global dust storm}.
\newblock {\em Geophys. Res. Lett.}

\bibitem[{Keating} et~al., 1998]{Keating1998}
{Keating}, G.~M., {Bougher}, S.~W., {Zurek}, R.~W., {Tolson}, R.~H., {Cancro},
  G.~J., {Noll}, S.~N., {Parker}, J.~S., {Schellenberg}, T.~J., {Shane}, R.~W.,
  {Wilkerson}, B.~L., {Murphy}, J.~R., {Hollingsworth}, J.~L., {Haberle},
  R.~M., {Joshi}, M., {Pearl}, J.~C., {Conrath}, B.~J., {Smith}, M.~D.,
  {Clancy}, R.~T., {Blanchard}, R.~C., {Wilmoth}, R.~G., {Rault}, D.~F.,
  {Martin}, T.~Z., {Lyons}, D.~T., {Esposito}, P.~B., {Johnston}, M.~D.,
  {Whetzel}, C.~W., {Justus}, C.~G., and {Babicke}, J.~M. (1998).
\newblock {The Structure of the Upper Atmosphere of Mars: In Situ Accelerometer
  Measurements from Mars Global Surveyor}.
\newblock {\em Science}, 279:1672.

\bibitem[{Kleinb{\"o}hl} et~al., 2017]{Kleinbohl2017}
{Kleinb{\"o}hl}, A., {Friedson}, A.~J., and {Schofield}, J.~T. (2017).
\newblock {Two-dimensional radiative transfer for the retrieval of limb
  emission measurements in the martian atmosphere}.
\newblock {\em J. Quant. Spectrosc. Radiat. Transf.}, 187:511--522.

\bibitem[{Kleinb{\"o}hl} et~al., 2013]{Kleinbohl2013}
{Kleinb{\"o}hl}, A., {John Wilson}, R., {Kass}, D., {Schofield}, J.~T., and
  {McCleese}, D.~J. (2013).
\newblock {The semidiurnal tide in the middle atmosphere of Mars}.
\newblock {\em Geophys. Res. Lett.}, 40(10):1952--1959.

\bibitem[{Kleinb{\"o}hl} et~al., 2011]{Kleinbohl2011}
{Kleinb{\"o}hl}, A., {Schofield}, J.~T., {Abdou}, W.~A., {Irwin}, P. G.~J., and
  {de Kok}, R.~J. (2011).
\newblock {A single-scattering approximation for infrared radiative transfer in
  limb geometry in the Martian atmosphere}.
\newblock {\em J. Quant. Spectrosc. Radiat. Transf.}, 112:1568--1580.

\bibitem[{Kleinb{\"o}hl} et~al., 2009]{Kleinbohl2009}
{Kleinb{\"o}hl}, A., {Schofield}, J.~T., {Kass}, D.~M., {Abdou}, W.~A.,
  {Backus}, C.~R., {Sen}, B., {Shirley}, J.~H., {Lawson}, W.~G., {Richardson},
  M.~I., {Taylor}, F.~W., {Teanby}, N.~A., and {McCleese}, D.~J. (2009).
\newblock {Mars Climate Sounder limb profile retrieval of atmospheric
  temperature, pressure, and dust and water ice opacity}.
\newblock {\em J. Geophys. Res. (Planets)}, 114(E10):E10006.

\bibitem[{Kleinb{\"o}hl} et~al., 2020]{Kleinbohl2020}
{Kleinb{\"o}hl}, A., {Spiga}, A., {Kass}, D.~M., {Shirley}, J.~H., {Millour},
  E., {Montabone}, L., and {Forget}, F. (2020).
\newblock {Diurnal Variations of Dust During the 2018 Global Dust Storm
  Observed by the Mars Climate Sounder}.
\newblock {\em J. Geophys. Res. (Planets)}, 125(1):e06115.

\bibitem[{Lee} et~al., 2009]{Lee2009}
{Lee}, C., {Lawson}, W.~G., {Richardson}, M.~I., {Heavens}, N.~G.,
  {Kleinb{\"o}hl}, A., {Banfield}, D., {McCleese}, D.~J., {Zurek}, R., {Kass},
  D., {Schofield}, J.~T., {Leovy}, C.~B., {Taylor}, F.~W., and {Toigo}, A.~D.
  (2009).
\newblock {Thermal tides in the Martian middle atmosphere as seen by the Mars
  Climate Sounder}.
\newblock {\em J. Geophys. Res. (Planets)}, 114(E3):E03005.

\bibitem[{Leovy}, 1981]{Leovy1981}
{Leovy}, C.~B. (1981).
\newblock {Observations of Martian tides over Two annual cycles.}
\newblock {\em J. Atmos. Sci.}, 38:30--39.

\bibitem[{Leovy} and {Zurek}, 1979]{LeovyZurek1979}
{Leovy}, C.~B. and {Zurek}, R.~W. (1979).
\newblock {Thermal tides and Martian dust storms: direct evidence for
  coupling.}
\newblock {\em J. Geophys. Res.}, 84:2956--2968.

\bibitem[{Leovy} et~al., 1973]{Leovy1973}
{Leovy}, C.~B., {Zurek}, R.~W., and {Pollack}, J.~B. (1973).
\newblock {Mechanisms for Mars dust storms.}
\newblock {\em J. Atmos. Sci.}, 30:749--762.

\bibitem[{Lo} et~al., 2015]{Lo2015}
{Lo}, D.~Y., {Yelle}, R.~V., {Schneider}, N.~M., {Jain}, S.~K., {Stewart}, A.
  I.~F., {England}, S.~L., {Deighan}, J.~I., {Stiepen}, A., {Evans}, J.~S.,
  {Stevens}, M.~H., {Chaffin}, M.~S., {Crismani}, M. M.~J., {McClintock},
  W.~E., {Clarke}, J.~T., {Holsclaw}, G.~M., {Lef{\`e}vre}, F., and {Jakosky},
  B.~M. (2015).
\newblock {Nonmigrating tides in the Martian atmosphere as observed by MAVEN
  IUVS}.
\newblock {\em Geophys. Res. Lett.}, 42(21):9057--9063.

\bibitem[{Maltagliati} et~al., 2011]{Maltagliati2011}
{Maltagliati}, L., {Titov}, D.~V., {Encrenaz}, T., {Melchiorri}, R., {Forget},
  F., {Keller}, H.~U., and {Bibring}, J.-P. (2011).
\newblock {Annual survey of water vapor behavior from the OMEGA mapping
  spectrometer onboard Mars Express}.
\newblock {\em Icarus}, 213(2):480--495.

\bibitem[{Mazarico} et~al., 2008]{Mazarico2008}
{Mazarico}, E., {Zuber}, M.~T., {Lemoine}, F.~G., and {Smith}, D.~E. (2008).
\newblock {Observation of atmospheric tides in the Martian exosphere using Mars
  Reconnaissance Orbiter radio tracking data}.
\newblock {\em Geophys. Res. Lett.}, 35(9):L09202.

\bibitem[{McCleese} et~al., 2007]{McCleese2007}
{McCleese}, D.~J., {Schofield}, J.~T., {Taylor}, F.~W., {Calcutt}, S.~B.,
  {Foote}, M.~C., {Kass}, D.~M., {Leovy}, C.~B., {Paige}, D.~A., {Read}, P.~L.,
  and {Zurek}, R.~W. (2007).
\newblock {Mars Climate Sounder: An investigation of thermal and water vapor
  structure, dust and condensate distributions in the atmosphere, and energy
  balance of the polar regions}.
\newblock {\em J. Geophys. Res. (Planets)}, 112(E5):E05S06.

\bibitem[{Montabone} et~al., 2015]{Montabone2015}
{Montabone}, L., {Forget}, F., {Millour}, E., {Wilson}, R.~J., {Lewis}, S.~R.,
  {Cantor}, B., {Kass}, D., {Kleinb{\"o}hl}, A., {Lemmon}, M.~T., {Smith},
  M.~D., and {Wolff}, M.~J. (2015).
\newblock {Eight-year climatology of dust optical depth on Mars}.
\newblock {\em Icarus}, 251:65--95.

\bibitem[{Moudden} and {Forbes}, 2008a]{MouddenForbes2008a}
{Moudden}, Y. and {Forbes}, J.~M. (2008a).
\newblock {Effects of vertically propagating thermal tides on the mean
  structure and dynamics of Mars' lower thermosphere}.
\newblock {\em Geophys. Res. Lett.}, 35(23):L23805.

\bibitem[{Moudden} and {Forbes}, 2008b]{MouddenForbes2008b}
{Moudden}, Y. and {Forbes}, J.~M. (2008b).
\newblock {Topographic connections with density waves in Mars' aerobraking
  regime}.
\newblock {\em J. Geophys. Res. (Planets)}, 113(E11):E11009.

\bibitem[{Moudden} and {Forbes}, 2010]{MouddenForbes2010}
{Moudden}, Y. and {Forbes}, J.~M. (2010).
\newblock {A new interpretation of Mars aerobraking variability: Planetary
  wave-tide interactions}.
\newblock {\em J. Geophys. Res. (Planets)}, 115(E9):E09005.

\bibitem[{Moudden} and {Forbes}, 2011]{MouddenForbes2011}
{Moudden}, Y. and {Forbes}, J.~M. (2011).
\newblock {Simulated planetary wave-tide interactions in the atmosphere of
  Mars}.
\newblock {\em J. Geophys. Res. (Planets)}, 116(E1):E01004.

\bibitem[{Moudden} and {Forbes}, 2014]{MouddenForbes2014}
{Moudden}, Y. and {Forbes}, J.~M. (2014).
\newblock {Insight into the seasonal asymmetry of nonmigrating tides on Mars}.
\newblock {\em Geophys. Res. Lett.}, 41(7):2631--2636.

\bibitem[{Shirley} et~al., 2020]{Shirley2020}
{Shirley}, J.~H., {Kleinb{\"o}hl}, A., {Kass}, D.~M., {Steele}, L.~J.,
  {Heavens}, N.~G., {Suzuki}, S., {Piqueux}, S., {Schofield}, J.~T., and
  {McCleese}, D.~J. (2020).
\newblock {Rapid Expansion and Evolution of a Regional Dust Storm in the
  Acidalia Corridor During the Initial Growth Phase of the Martian Global Dust
  Storm of 2018}.
\newblock {\em Geophys. Res. Lett.}, 47(9):e84317.

\bibitem[{Smith}, 2002]{Smith2002b}
{Smith}, M.~D. (2002).
\newblock {The annual cycle of water vapor on Mars as observed by the Thermal
  Emission Spectrometer}.
\newblock {\em J. Geophys. Res. (Planets)}, 107(E11):5115.

\bibitem[{Smith}, 2004]{Smith2004}
{Smith}, M.~D. (2004).
\newblock {Interannual variability in TES atmospheric observations of Mars
  during 1999-2003}.
\newblock {\em Icarus}, 167(1):148--165.

\bibitem[{Steele} et~al., 2014]{Steele2014}
{Steele}, L.~J., {Lewis}, S.~R., {Patel}, M.~R., {Montmessin}, F., {Forget},
  F., and {Smith}, M.~D. (2014).
\newblock {The seasonal cycle of water vapour on Mars from assimilation of
  Thermal Emission Spectrometer data}.
\newblock {\em Icarus}, 237:97--115.

\bibitem[{Stevens} et~al., 2017]{Stevens2017}
{Stevens}, M.~H., {Siskind}, D.~E., {Evans}, J.~S., {Jain}, S.~K., {Schneider},
  N.~M., {Deighan}, J., {Stewart}, A.~I.~F., {Crismani}, M., {Stiepen}, A.,
  {Chaffin}, M.~S., {McClintock}, W.~E., {Holsclaw}, G.~M., {Lef{\`e}vre}, F.,
  {Lo}, D.~Y., {Clarke}, J.~T., {Montmessin}, F., and {Jakosky}, B.~M. (2017).
\newblock {Martian mesospheric cloud observations by IUVS on MAVEN: Thermal
  tides coupled to the upper atmosphere}.
\newblock {\em Geophys. Res. Lett.}, 44(10):4709--4715.

\bibitem[{Trokhimovskiy} et~al., 2015]{Trokhimovskiy2015}
{Trokhimovskiy}, A., {Fedorova}, A., {Korablev}, O., {Montmessin}, F.,
  {Bertaux}, J.-L., {Rodin}, A., and {Smith}, M.~D. (2015).
\newblock {Mars' water vapor mapping by the SPICAM IR spectrometer: Five
  martian years of observations}.
\newblock {\em Icarus}, 251:50--64.

\bibitem[{Vi{\'u}dez-Moreiras} et~al., 2019]{Viudez2019}
{Vi{\'u}dez-Moreiras}, D., {Newman}, C.~E., {de la Torre}, M., {Mart{\'\i}nez},
  G., {Guzewich}, S., {Lemmon}, M., {Pla-Garc{\'\i}a}, J., {Smith}, M.~D.,
  {Harri}, A.~M., {Genzer}, M., {Vicente-Retortillo}, A., {Lepinette}, A.,
  {Rodriguez-Manfredi}, J.~A., {Vasavada}, A.~R., and {G{\'o}mez-Elvira}, J.
  (2019).
\newblock {Effects of the MY34/2018 Global Dust Storm as Measured by MSL REMS
  in Gale Crater}.
\newblock {\em J. Geophys. Res. (Planets)}, 124(7):1899--1912.

\bibitem[{Wang} and {Richardson}, 2015]{WangRichardson2015}
{Wang}, H. and {Richardson}, M.~I. (2015).
\newblock {The origin, evolution, and trajectory of large dust storms on Mars
  during Mars years 24-30 (1999-2011)}.
\newblock {\em Icarus}, 251:112--127.

\bibitem[{Wang} et~al., 2006]{Wang2006}
{Wang}, L., {Fritts}, D.~C., and {Tolson}, R.~H. (2006).
\newblock {Nonmigrating tides inferred from the Mars Odyssey and Mars Global
  Surveyor aerobraking data}.
\newblock {\em Geophys. Res. Lett.}, 33(23):L23201.

\bibitem[{Wilson}, 2000]{Wilson2000}
{Wilson}, R.~J. (2000).
\newblock {Evidence for diurnal period Kelvin waves in the Martian atmosphere
  from Mars Global Surveyor TES data}.
\newblock {\em Geophys. Res. Lett.}, 27(23):3889--3892.

\bibitem[{Wilson}, 2002]{Wilson2002}
{Wilson}, R.~J. (2002).
\newblock {Evidence for nonmigrating thermal tides in the Mars upper atmosphere
  from the Mars Global Surveyor Accelerometer Experiment}.
\newblock {\em Geophys. Res. Lett.}, 29(7):1120.

\bibitem[{Wilson} and {Guzewich}, 2014]{Wilson2014}
{Wilson}, R.~J. and {Guzewich}, S.~D. (2014).
\newblock {Influence of water ice clouds on nighttime tropical temperature
  structure as seen by the Mars Climate Sounder}.
\newblock {\em Geophys. Res. Lett.}, 41(10):3375--3381.

\bibitem[{Wilson} and {Hamilton}, 1996]{WilsonHamilton1996}
{Wilson}, R.~J. and {Hamilton}, K. (1996).
\newblock {Comprehensive model simulation of thermal tides in the Martian
  atmosphere.}
\newblock {\em J. Atmos. Sci.}, 53(9):1290--1326.

\bibitem[{Wilson} and {Richardson}, 2000]{WilsonRichardson2000}
{Wilson}, R.~J. and {Richardson}, M.~I. (2000).
\newblock {The Martian atmosphere during the Viking mission. I. Infrared
  measurements of atmospheric temperatures revisited.}
\newblock {\em Icarus}, 145(2):555--579.

\bibitem[{Withers} et~al., 2003]{Withers2003}
{Withers}, P., {Bougher}, S.~W., and {Keating}, G.~M. (2003).
\newblock {The effects of topographically-controlled thermal tides in the
  martian upper atmosphere as seen by the MGS accelerometer}.
\newblock {\em Icarus}, 164(1):14--32.

\bibitem[{Withers} et~al., 2011]{Withers2011}
{Withers}, P., {Pratt}, R., {Bertaux}, J.~L., and {Montmessin}, F. (2011).
\newblock {Observations of thermal tides in the middle atmosphere of Mars by
  the SPICAM instrument}.
\newblock {\em J. Geophys. Res. (Planets)}, 116(E11):E11005.

\bibitem[{Wolkenberg} et~al., 2011]{Wolkenberg2011}
{Wolkenberg}, P., {Smith}, M.~D., {Formisano}, V., and {Sindoni}, G. (2011).
\newblock {Comparison of PFS and TES observations of temperature and water
  vapor in the martian atmosphere}.
\newblock {\em Icarus}, 215(2):628--638.

\bibitem[{Wu} et~al., 2015]{Wu2015}
{Wu}, Z., {Li}, T., and {Dou}, X. (2015).
\newblock {Seasonal variation of Martian middle atmosphere tides observed by
  the Mars Climate Sounder}.
\newblock {\em J. Geophys. Res. (Planets)}, 120(12):2206--2223.

\bibitem[{Wu} et~al., 2017]{Wu2017}
{Wu}, Z., {Li}, T., and {Dou}, X. (2017).
\newblock {What causes seasonal variation of migrating diurnal tide observed by
  the Mars Climate Sounder?}
\newblock {\em J. Geophys. Res. (Planets)}, 122(6):1227--1242.

\bibitem[{Zhao} et~al., 2015]{Zhao2015}
{Zhao}, Y., {Greybush}, S.~J., {Wilson}, R.~J., {Hoffman}, R.~N., and {Kalnay},
  E. (2015).
\newblock {Impact of assimilation window length on diurnal features in a Mars
  atmospheric analysis}.
\newblock {\em Tellus A}, 67(1):26042.

\bibitem[{Zurek}, 1976]{Zurek1976}
{Zurek}, R.~W. (1976).
\newblock {Diurnal tide in the Martian atmosphere.}
\newblock {\em J. Atmos. Sci.}, 33:321--337.

\bibitem[{Zurek}, 1980]{Zurek1980}
{Zurek}, R.~W. (1980).
\newblock {Surface pressure response to elevated tidal heating sources:
  comparison of Earth and Mars.}
\newblock {\em J. Atmos. Sci.}, 37:1132--1136.

\bibitem[{Zurek}, 1988]{Zurek1988}
{Zurek}, R.~W. (1988).
\newblock {Free and forced modes in the Martian atmosphere.}
\newblock {\em J. Geophys. Res.}, 93:9452--9462.

\bibitem[{Zurek} and {Leovy}, 1981]{Zurek1981}
{Zurek}, R.~W. and {Leovy}, C.~B. (1981).
\newblock {Thermal Tides in the Dusty Martian Atmosphere: A Verification of
  Theory}.
\newblock {\em Science}, 213(4506):437--439.

\end{thebibliography}

\end{document}